\documentclass[12pt,a4paper]{article}
\usepackage[usenames,dvipsnames]{pstricks} % latexdraw
\usepackage{afterpage}
\usepackage{epsfig}
\usepackage{pst-grad} % For gradients
\usepackage{pst-plot} % For axes
\usepackage[a4paper]{geometry}
\usepackage{float}
\usepackage[stable]{footmisc}
\usepackage[utf8x]{inputenc}
\usepackage{a4wide}
\usepackage{amsmath,amssymb,amstext,amsthm,amssymb}
\usepackage{amsfonts}
\usepackage{framed}
\usepackage{graphicx}
\usepackage{color}
\usepackage{bbold}
\usepackage{bbm}
\usepackage{rotating} %% rotating figures
\usepackage{slashed} %%feynman slash
\usepackage{cancel} %% diagonal line through word
\usepackage{braket} %% bra ket
\usepackage{amstext}
\usepackage{array}
\usepackage{setspace}
\usepackage{hyperref}
\usepackage{overpic}
\usepackage{bbm}
\usepackage{cite}
\usepackage{lipsum}
\newcommand{\del}[0]{\partial}
\let\baraccent=\=
\renewcommand{\=}[1]{\stackrel{#1}{=}}
\newcommand{\id}[0]{\mathbb{1}}
\newcommand{\Tr}[1]{\text{Tr}\left( #1 \right)}
\title{Towards de Sitter from 10D}
\begin{document}
	\rightline{DESY 17-112}\vspace*{3ex}
	\title{Towards de Sitter from 10D}
	
	\pagenumbering{gobble}
	\begin{center}
		{\huge \textbf{Towards de Sitter from 10D}}\\
		\vspace{1.5cm}
		{\large Jakob Moritz$^1$, Ander Retolaza$^2$ and Alexander Westphal$^3$}\\
		\vspace{0.5cm}
		\textit{
			 Deutsches Elektronen-Synchrotron, DESY, Notkestra\ss e 85, 22607 Hamburg, Germany}\\
		\vspace{1.5cm}
		\textbf{Abstract}\\
	\end{center}
	\vspace{0.5cm}
Using a 10D lift of non-perturbative volume stabilization in type IIB string theory we study the limitations for obtaining de Sitter vacua. Based on this we find that the simplest KKLT vacua with a single K\"ahler modulus stabilized by a gaugino condensate cannot be uplifted to de Sitter. Rather, the uplift \textit{flattens out} due to stronger back-reaction on the volume modulus than has  previously been anticipated, resulting in vacua which are meta-stable and SUSY breaking, but that are always $AdS$. 
However, we also show that setups such as racetrack stabilization can avoid this issue. In these models it is possible to obtain supersymmetric $AdS$ vacua with a cosmological constant that can be tuned to zero while retaining finite moduli stabilization. In this regime, it seems that de Sitter uplifts are possible with negligible backreaction on the internal volume. We exhibit this behavior also from the  $10D$ perspective.

	\vspace*{10ex}
	\noindent July 26, 2017\vspace*{5ex}\\
	
	\noindent $^1$\href{mailto:jakob.moritz@desy.de}{jakob.moritz@desy.de}\\
	\noindent$^2$\href{mailto:ander.retolaza@desy.de}{ander.retolaza@desy.de}\\
	\noindent$^3$\href{mailto:alexander.westphal@desy.de}{alexander.westphal@desy.de}

	\newpage
	\pagenumbering{arabic}
	
	\section{Introduction}
\label{intro}

%%%%%%%%%%%%%%%%%%%%%%%%%%%%%%%%%%%%%%%%%%%%%%%%%%%%%%%%%%%%%%%%%%%%%%

The cosmological constant (c.c.) problem stands as our perhaps most drastic disagreement between theory and observation. When applying quantum field theory (QFT) to various stages of symmetry breaking and to the Einstein field theory limit of gravity given by general relativity (GR), we arrive at a prediction which is off by some 120 orders of magnitude from the value of the c.c. inferred from, for example, precision observations of the cosmic microwave background (CMB).

Attempting to resolve this tension is difficult given a no-go theorem by Weinberg~\cite{Weinberg:1988cp} which essentially states that dynamic adjustment mechanisms based on a fully local field theory only must fail. This leaves us with few options. 

One possibility is to give up locality, motivated by the fact that a true c.c. needs the full 4-volume of the universe for a  proper detection, which makes it borderline non-local. Recent work has shown that a combination of protective scale and shift symmetries with the introduction of certain effectively global variables encoded via 4-form field strengths can `sequester' the QFT vacuum energy from its gravitating effects~\cite{Kaloper:2013zca,Kaloper:2015jra,Kaloper:2016jsd}. 

Another possibility, already foreseen by Linde, Weinberg and others~\cite{Linde:1984ir,Linde:1986dq,Hawking:1987en,Weinberg:1987dv,Barrow:1988yia}, is to replace the  QFT-limit of GR with a full candidate theory of quantum gravity, and assume that it will render the quantum vacuum energy finite and generically large, but produce an enormous number of different solutions (vacua) with differing finite c.c.'s. If we couple this with a cosmological population mechanism such as eternal inflation~\cite{Vilenkin:1983xq,Linde:1986fd}, then a form of weak anthropic argument can explain the observed c.c. value as natural given certain observed features such as structure formation, and provided the number of different vacua is larger than ${\cal O}(\Lambda^{-4})={\cal O}(10^{120})$ (and the distribution of vacuum energies shows no strong features).

Crucial for this line of argument is having in hand such a candidate quantum gravity with an enormous `landscape' of discrete vacua with varying c.c. values~\cite{Susskind:2003kw}.

One of the first serious and still one of the best candidates for this task is the setup by Kachru, Kallosh, Linde and Trivedi (KKLT)~\cite{Kachru:2003aw} which uses string theory as a candidate quantum gravity to realize a version of the Bousso-Polchinski proposal, a toy model of the landscape solution to the c.c. problem.

KKLT propose to generate de Sitter vacua in string theory in a three-step process: 
\begin{itemize}
	\item Step 1: Stabilize complex structure moduli and the axio-dilaton perturbatively using three-form fluxes~\cite{Giddings:2001yu,Dasgupta:1999ss}.
	\item Step 2: Stabilize K\"ahler moduli using non-perturbative effects. The result is a supersymmetric $AdS$ vacuum~\cite{Kachru:2003aw}.
	\item Step 3: Break supersymmetry and lift to de Sitter space by including an anti-brane at the bottom of a warped throat~\cite{Kachru:2002gs,Kachru:2003aw}.
\end{itemize}

In looking at these three steps we observe the following situation. The 4D effective supergravity description of step 1 is derived via dimensional reduction from the warped CY compactifications with ISD 3-form fluxes, and its critical points match with solutions of the full $10D$ equations of motion, the $10D$ Einstein equations and Bianchi identities supplemented by $O3/O7$ planes and $D3/D7$ branes as local sources. The volume modulus $\rho$ is a flat direction at this level~\cite{Giddings:2001yu}.

The $4D$ effective description of step 2 is based on the $4D$ effective theory of gaugino condensation~\cite{Veneziano:1982ah,Ferrara:1982qs,Affleck:1983mk,Dine:1985rz}. However, the presence of unbroken supersymmetry in the ensuing AdS vacuum of $\rho$ as well as the suppression of the $\rho$ mass scale with the control parameter $|W_0|$ given by the fluxes provides justification for the choice of the form of the 4D EFT of the $\rho$ modulus, as it is more or less dictated by holomorphy of the superpotential, supersymmetry of the setup, non-perturbative gauge dynamics, and the relation between the holomorphic gauge kinetic function of the D7-brane gauge theory and the 4-cycle volume modulus chiral multiplet $f_{D7}(\Sigma_4^i)\sim \rho_i$. Moreover, all scales of the volume stabilization process lie far below those of the flux stabilization of step 1~\cite{Kachru:2003aw}.

There is by now very strong evidence that step 3, inserting in particular a single anti-D3-brane at the bottom of the warped throat of step 1 + step 2, is still a well-controlled meta-stable solution of the full system with spontaneously broken supersymmetry. Its non-perturbative instability arises from the Kachru-Pearson-Verlinde (KPV) flux-brane annihilation process~\cite{Kachru:2002gs,Freivogel:2008wm}. Describing this system beyond the probe approximation for the anti-D3-brane~\cite{Blaback:2012nf,Bena:2012tx,Bena:2012vz,Bena:2012ek,Bena:2013hr,Junghans:2014wda,Bena:2014jaa,Danielsson:2014yga,Bena:2015kia,Cohen-Maldonado:2015ssa,Hartnett:2015oda,Danielsson:2015eqa} while respecting a well-defined EFT description of the anti-D3 backreaction onto the warped throat geometry, with correct matching onto the string scale physics~\cite{Michel:2014lva,Polchinski:2015bea} provides clear arguments that the KPV process is the microscopic non-perturbative instability of the setup with all perturbative directions (the anti-D3 position moduli) frozen.

However, we observe a more subtle peculiarity of this uplift effect in the KKLT setting. Namely, the amount of uplift necessary to reach de Sitter is tied to the mass scale of the $\rho$ modulus from its stabilization in step 2. To reach de Sitter we require validity of a particular form of the 4D EFT. Modifications to the EFT that are compatible with our current knowledge of the supersymmetric moduli stabilization as well as the classical $\overline{D3}$-brane potential can lead to substantial backreaction effects on the volume modulus. In the extreme case these modifications could lead to a \textit{flattened} uplift that never reaches a positive vacuum potential. `Flattening' means that the backreacted uplift turns out to be smaller than the estimate without backreaction which scales linear with the anti-brane tension.

%Once backreaction is included, we find that the backreacted amount of uplifting energy \emph{flattens} -- it turns out to be smaller than the no-backreaction estimate which scales linear with the anti-brane tension. In the extreme case these modification could lead to a \textit{flattened} uplift that never reaches a positive vacuum potential.

We note here, that such flattening effects from moduli backreaction were observed before in a related context: namely, if we replace the control parameter of the uplifting vacuum energy, the anti-brane tension, by the potential energy of a slow-rolling scalar field driving inflation, the same backreaction effects lead to \emph{flattened scalar potentials} for inflation~\cite{Dong:2010in}, in complete analogy with the \emph{uplift flattening} we observe here.

Indeed such modifications to the EFT can be argued to arise through the same effects that lead to the lifting of the moduli space of $D3$-brane position moduli by non-perturbative effects. However, the very same reasoning suggests that if the racetrack stabilization of Kallosh and Linde \cite{Kallosh:2004yh} is assumed, de Sitter uplifts can be realized because the $AdS$ depth of the supersymmetric potential can be parametrically decoupled from the mass scale of the K\"ahler modulus.

Consequently, the questions we would like to ask and partially answer in this paper in the course of sections~\ref{KKLT} through~\ref{10D} are as follows. Is the volume stabilization of the original KKLT proposal rigid enough to allow for a de Sitter uplift or is racetrack stabilization a minimal requirement? If the latter is true, can we verify the stabilization mechanism also from a $10D$ perspective?

We find indeed that back-reaction effects play a  substantially stronger role than has previously been anticipated. The \emph{uplift flattening} we observe is strong enough that uplifts reaching de Sitter are not possible using the simplest version of KKLT. Instead the solutions for volume stabilization with a \emph{single} gaugino condensate are meta-stable SUSY breaking AdS vacua. However, assuming certain properties of the compactification manifold, we are able to give a $10D$ picture of \emph{racetrack stabilization} that strongly indicates the possibility to uplift to de Sitter in this case. 

We use these 4D insights to build a modified single-condensate moduli stabilization scheme by including $\alpha '$ corrections. This allows to decouple the mass scale of the K\"ahler modulus from the $AdS$ depth, although not parametrically.

We then discuss our findings and conclude in section~\ref{Concl}.

%\subsection{ Conventions  }\label{conventions}

%%%%%%%%%%%%%%%%%%%%%%%%%%%%%%%%%%%%%%%%%%%%%%%%%%%%%%%%%%%%%%%%%%%%%%

%Throughout this paper we will always consider spacetimes with  $D$ dimensions with metric signature $(-,+,...,+)$. These spacetimes will always have 4 non-compact directions as well as $d=D-4$ compact ones. Since we are specially interested in compactifications of string theory, we will often have $D=10$, but other situations with $D=6$ will also be used in toy models in order to provide an intuitive warm-up picture. Capital roman indices $M,N,...$ run over all dimensions $0,...,D-1$ while greek indices $\mu,\nu,...$ run over the non-compact directions $0,...,3$ and lower-case roman indices $m,n,...$ run over internal directions $4,...,D-1$.

%Furthermore we define the Hodge-star such that 
%\begin{equation}
%||F_p||^2\equiv\int F_p\wedge *F_p\equiv\int d^D x\sqrt{-g}|F_p|^2\equiv \int d^D x\sqrt{-g}\frac{1}{p!}F_{O_1\cdots O_p}F^{O_1\cdots O_p}\, ,
%\end{equation} for a $p$-form $F_p$. This implies that if a $p$-form has a leg in the time direction, $|F_p|^2<0$.\par

%%%%%%%%%%%%%%%%%%%%%%%%%%%%%%%%%%%%%%%%%%%%%%%%%%%%%%%%%%%%%%%%%%%%%%
\section{Moduli Stabilization and de Sitter uplifts in type IIB String Theory: The $4D$ perspective}\label{KKLT}
%%%%%%%%%%%%%%%%%%%%%%%%%%%%%%%%%%%%%%%%%%%%%%%%%%%%%%%%%%%%%%%%%%%%%%

The best understood constructions of de Sitter vacua in type IIB string theory start from Calabi-Yau (CY) compactifications to $4D$ Minkowski and orientifolds thereof. These compactifications  come with many massless moduli which need to be stabilized in order to make them phenomenologically interesting and typically require some extra structure/ingredient to give rise to  4D  de Sitter. This program of `moduli stabilization' and de Sitter uplift usually is carried out in three steps:
\begin{itemize}
	\item Stabilization of complex structure moduli and the axio-dilaton at tree-level by introduction of three-form fluxes \cite{Giddings:2001yu}.
	\item Stabilization of K\"ahler moduli. Different approaches exist to stabilize them, probably the most common ones being the following: in the so-called KKLT approach \cite{Kachru:2003aw},  one  stabilizes them using non-perturbative effects,   typically leading  to a  supersymmetric anti de Sitter (AdS) geometry in 4D. In the large volume scenario \cite{Balasubramanian:2005zx}, instead, they are stabilized using an interplay of $\alpha'$ corrections with non-perturbative corrections, such that the 4D vacuum is also AdS but non-supersymmetric. In what follows we will mainly focus on the former approach, the only exception being section \ref{sec:alphaprime}, where we will combine $\alpha '$ corrections with a gaugino condensate to stabilize the K\"{a}hler modulus.
	\item The final step consists of uplifting the 4D geometry to de Sitter space. Many proposals exist to achieve this goal (see e.g.  \cite{Burgess:2003ic,Saltman:2004jh,Westphal:2006tn,Cicoli:2012fh,Louis:2012nb,Rummel:2014raa,Cicoli:2015ylx,Braun:2015pza,Retolaza:2015nvh,Gallego:2017dvd}), the most common one being  the inclusion of a supersymmetry breaking source, most prominently an $\overline{D3}$-brane, at the bottom of a warped throat \cite{Kachru:2003aw}.
\end{itemize}
These steps are usually followed through entirely within the framework of $4D$ supergravity, the low energy effective field theory describing the  4D fields after the compactification.   In this section we will quickly review some of its aspects.\par 
Before doing so, let us give the main point of this section right away: the first two steps can be followed with reasonable amount of control purely within the $4D$ effective field theory. However, in many models of moduli stabilization \cite{Kachru:2003aw} the uplift can not be parametrically decoupled from K\"ahler moduli stabilization. In these cases, although we believe that there is nothing wrong with the use of $4D$ supergravity as a matter of principle, the question whether or not a proposed uplift can be successful heavily depends on the detailed moduli dependence of the uplift ingredient. We will argue that at least for $\overline{D3}$-brane uplifts the required knowledge about the moduli dependence is hard to acquire from $4D$ effective field theory reasoning alone. In other words, guessing the correct effective field theory is not as obvious as it might seem.

%%%%%%%%%%%%%%%%%%%%%%%%%%%%%%%%%%%%%%%%%%%%%%%%%%%%%%%%%%%%%%%%%%%%%%
\subsection{Complex Structure Moduli Stabilization}\label{KKLT:csstab}
%%%%%%%%%%%%%%%%%%%%%%%%%%%%%%%%%%%%%%%%%%%%%%%%%%%%%%%%%%%%%%%%%%%%%%

The effective $4D$ supergravity that governs the light moduli of type IIB $O3/O7$ CY-orientifolds is well known. For  a single K\"ahler modulus (i.e. $h_{1,1}^+=1$) the K\"ahler potential is given by  
\begin{equation}
\mathcal{K}(\rho,z_{\alpha},\tau)=-3\ln (-i(\rho-\bar{\rho}))-\ln(-i(\tau-\bar{\tau}))-\ln(-i\int_{M_6}\Omega\wedge \overline{\Omega})\, ,
\end{equation}
to leading order in the $\alpha'$ expansion. Since we are considering a unique K\"{a}hler modulus $\rho$, this is the (complexified) volume modulus ($\text{Im}(\rho)=\mathcal{V}^{2/3}$). Also, $\tau=C+ie^{-\phi}$ is the axio-dilaton and $z_{\alpha}$ are the $h_{2,1}^-$ complex structure moduli that appear in the K\"ahler potential implicitly via the holomorphic 3-form  $\Omega=\Omega(z_{\alpha})$.\par
Dilute three-form fluxes (such that warping effects are negligible) generate a superpotential, the Gukov-Vafa-Witten (GVW) superpotential \cite{Gukov:1999ya}
\begin{equation}
W_{GVW}=\sqrt{\frac{2}{\pi}}\int_{M_6} G_3\wedge \Omega\, ,
\end{equation}
where  $G_3=F_3-\tau H_3$ is the complexified 3-form flux. This superpotential depends holomorphically on the axio-dilaton and the complex structure moduli. It was derived by comparing the tension $\mathcal{T}^{DW}$ of BPS domain walls that separate flux vacua with different three-form fluxes (i.e. $D5$/$NS5$ branes that wrap internal three-cycles) with the change in superpotential $\Delta W$ across the domain wall and demand the standard $4D$ SUSY relation
\begin{equation}
\mathcal{T}^{DW}=2e^{\mathcal{K}/2}|\Delta W|\, .
\end{equation}
The SUSY conditions $D_iW=0$ (where $i$ runs over the complex structure moduli and the axio-dilaton) require that $G_3$ is of Hodge type $(2,1)\oplus (0,3)$. This should be read as an equation for the complex structure moduli and the axio-dilaton, which obtain masses of the order
\begin{equation}
m_{CS}\sim \frac{\alpha'}{R^3}\  ,
\end{equation}
where $R$ is a typical scale of the internal geometry. Thus, at large volume they are much lighter than the Kaluza-Klein (KK) modes of the internal geometry.
If the superpotential is entirely determined by the GVW superpotential, which does not depend on the K\"{a}hler modulus $\rho$, the $4D$  potential vanishes at the minimum of the complex structure moduli and the axio-dilaton due to the \textit{no-scale} relation  $ \mathcal{K}^{\rho\bar{\rho}}\del_{\rho}\mathcal{K}\del_{\bar{\rho}} \mathcal{K}=3$. This means that the volume modulus remains massless and the final SUSY condition
\begin{equation}
0=D_{\rho}W=(\del_{\rho}\mathcal{K})W
\end{equation}
cannot be solved at finite volume unless the non-generic condition $W=0$ happens to be met, i.e. $G_3$ is of Hodge type $(2,1)$.\par
The upshot is that for generic fluxes all but the single K\"ahler modulus are stabilized and SUSY is broken. Hence, at low energies one is left with a single modulus $\rho$ and a constant superpotential $W=W_0$.

%%%%%%%%%%%%%%%%%%%%%%%%%%%%%%%%%%%%%%%%%%%%%%%%%%%%%%%%%%%%%%%%%%%%%%
\subsection{K\"ahler Moduli Stabilization: KKLT}\label{KKLT:Kstab}
%%%%%%%%%%%%%%%%%%%%%%%%%%%%%%%%%%%%%%%%%%%%%%%%%%%%%%%%%%%%%%%%%%%%%%

In order to stabilize the volume modulus KKLT  argued for a $\rho$-dependent correction of the superpotential \cite{Kachru:2003aw}. Because the superpotential must depend holomorphically on the volume modulus and the real part of $\rho$ enjoys a continuous shift symmetry to all orders in perturbation theory, the only possible corrections to the superpotential are non-perturbative
\begin{equation}
\delta W_{\text{np}} =\sum_n A_n e^{ia_n\rho} \quad,
\end{equation}
and break the continuous shift symmetry of $\text{Re}(\rho)$ to a discrete one.
There are two kinds of non-perturbative effects that can generate the desired exponential terms:
\begin{enumerate}
	\item Euclidean $D3$ branes  wrapping  internal $4$-cycles  \cite{Witten:1996bn}.
	\item Gaugino condensation on a stack of $N$ $D7$-branes \cite{Veneziano:1982ah,Ferrara:1982qs,Affleck:1983mk,Dine:1985rz}.
\end{enumerate}
In what follows we will focus  on (b): a stack of $N$ $D7$-branes wrapping a \textit{rigid} $4$-cycle $\Sigma$, i.e. the position moduli of the $D7$'s are massive. Assuming that no further branes intersect the $7$-brane stack (i.e. light flavors are absent) the low energy effective theory is  $SU(N)$ SYM. At low energies this theory confines and develops a non-perturbative superpotential via gaugino condensation,
\begin{equation}
\langle \lambda\lambda \rangle=-32\pi^2\Lambda^3\, , \quad \Rightarrow\quad 
\delta W_{\text{np}}=-\frac{N}{32\pi^2}\langle \lambda\lambda\rangle=N\Lambda^3\, ,
\end{equation}
where $\Lambda$ is the dynamical scale of the gauge theory which determines the perturbative running of the holomorphic gauge coupling $\tau_{YM}=\frac{\Theta_{YM}}{2\pi}+i \frac{4\pi}{g_{YM}^2}$  via 
\begin{equation}
\tau_{YM}=\frac{3N}{2\pi i}\ln\left(\frac{\Lambda}{\mu}\right),\quad \text{i.e.} \quad \Lambda=\mu e^{\frac{2\pi i}{3N}\tau_{YM}}\, .
\end{equation}
By dimensionally reducing the $D7$ brane action to $4D$ one can infer that classically the holomorphic gauge coupling is identified with the K\"ahler modulus $\rho$.
In the quantum theory this equation should be read as the renormalization condition that matches the low energy effective theory to its UV completion. In other words
\begin{equation}
\Lambda=\mu_0 e^{\frac{2\pi i}{3N}\rho}\, ,
\end{equation}
where $\mu_0$ is a high scale. In summary, there is a non-perturbative correction of the superpotential
\begin{equation}
\delta W_{\text{np}}(\rho)=N\mu_0^3 e^{\frac{2\pi i}{N}\rho}\, ,
\end{equation}
that we may associate to a non-vanishing expectation value of the gaugino condensate $\langle \lambda\lambda \rangle$.
Combining this with the constant flux superpotential $W_0$, at energies below the non-perturbative scale of the gauge theory the K\"ahler and superpotential reads
\begin{equation}
\mathcal{K}=-3\ln(-i(\rho-\bar\rho))\quad,\quad W=W_0+Ae^{ia\rho}\, .
\end{equation}
Now the F-term equation $D_{\rho}W=0$ is solved by
\begin{equation}
\label{W0_relationto_ll}
W_0=-\left(1+\frac{2a \text{Im}(\rho)}{3}\right)Ae^{ia\rho}\, ,
\end{equation}
and at the minimum of the potential the value of the vacuum energy is negative,
\begin{equation}
\label{4dPotential_at_min}
V=-3e^{\mathcal{K}}|W|^2=-\frac{a^2|A|^2}{12\text{Im}\rho}e^{-2a\,\text{Im}\rho}<0\, .
\end{equation}
Control over the instanton expansion requires $a\rho>0$ and control over the $\alpha'$ expansion of string theory requires $\rho\gg 0$. Both can be satisfied if due to a sufficient tuning of the flux configuration $|W_0|\ll 1$.\par 
An essential feature of this stabilization mechanism is that the value of the cosmological constant and the mass of the volume modulus are related via
\begin{equation}
|V_{AdS} | \sim M_{\rm P}^2 m_{\rho}^2\, .
\end{equation}

%%%%%%%%%%%%%%%%%%%%%%%%%%%%%%%%%%%%%%%%%%%%%%%%%%%%%%%%%%%%%%%%%%%%%%
\subsection{The uplift}\label{KKLT:uplift}
%%%%%%%%%%%%%%%%%%%%%%%%%%%%%%%%%%%%%%%%%%%%%%%%%%%%%%%%%%%%%%%%%%%%%%

The perhaps best understood way to uplift SUSY $AdS$-vacua of type IIB string theory to de Sitter is by the inclusion of an anti-D3 brane at the bottom of a warped throat. This way of uplifting is attractive because in ISD flux backgrounds the anti-brane is driven to strongly warped regions where its associated potential energy is naturally small \cite{Kachru:2002gs}. Moreover such SUSY-breaking vacua are known to be connected to the SUSY ISD vacua via the non-perturbative KPV transition \cite{Kachru:2002gs}. Arguably, no perturbative decay channel exists \cite{Polchinski:2015bea}. In this paper we assume perturbative stability of such a configuration. \par 
Because the $10D$ warped throat supergravity solution is dual to the Klebanov-Strassler (KS) gauge theory \cite{Klebanov:2000hb}, one should be able to equivalently describe the anti-brane as a state of the KS gauge theory that breaks supersymmetry spontaneously rather than explicitly \cite{Polchinski:2015bea}. If this is the case, it has been argued that at very low energies the only degrees of freedom are the nilpotent goldstino multiplet $S$ (nilpotency means $S^2=0$, for constrained superfields see e.g.~\cite{Komargodski:2009rz}) and the volume modulus $\rho$ \cite{Volkov:1973ix,Choi:2005ge,Lebedev:2006qq,Brummer:2006dg,Antoniadis:2014oya,Ferrara:2014kva,Kallosh:2014wsa,Bergshoeff:2015jxa,Kallosh:2015nia}. The proposals for the  K\"ahler- and the superpotential were
\begin{equation}
K=-3\ln (-i(\rho-\bar{\rho})-S\bar{S})\,\quad \text{and}\quad W=W_0+Ae^{ia\rho}+e^{2\mathcal{A}_{0}}\mu^2S\, . \label{eq:supo-KKLT}
\end{equation}
Here, $\exp(2\mathcal{A}_{0})\mu^2$ parametrizes the strength of supersymmetry breaking with $\exp(\mathcal{A}_{0})$ being the warp factor at the tip of throat, while $\mu$ is related to the unwarped tension of the anti-$D3$ brane as $|\mu|^4\sim T_3$.\par 

In deriving the scalar potential one should treat $S$ as a usual chiral multiplet and in the end set $S=0$. For real parameters $W_0$ and $A$, the scalar potential reads
\begin{equation}
\label{KKLTupliftpotential}
V(\rho)=\frac{aAe^{-a\,\text{Im}(\rho)}}{6\,\text{Im}(\rho)^2}\left[Ae^{-a\,\text{Im}(\rho)}(a\,\text{Im}(\rho)+3)+3W_0\cos\left(a\,\text{Re}(\rho)\right)\right]+e^{4\mathcal{A}_{0}}\,\frac{|\mu|^4}{12\,\text{Im}(\rho)^2}\, .
\end{equation}
The expectation of this being the correct way to describe the anti-brane state comes from the fact that in the limit of vanishing non-perturbative stabilization $A\longrightarrow 0$ one recovers the known runaway potential that is easily read off of the anti-brane DBI+CS action, while in the limit $\mu\longrightarrow 0$ one recovers the supersymmetric KKLT potential. The corresponding potential is simply the sum of the (would-be) runaway $\overline{D3}$ potential and the (would-be) supersymmetric KKLT potential.  Tuning the   tension contribution   of the anti-brane (by tuning the fluxes that determine the warp factor $e^{\mathcal{A}_{0}}$) one can obtain de Sitter vacua with tunably small cosmological constant \cite{Kachru:2003aw}. Interestingly, within this description of the $\overline{D3}$-induced uplift, the potential energy of the $\overline{D3}$ adds on top of the negative potential at the supersymmetric $AdS$ minimum to good approximation until a maximum uplift of about $\delta V \sim 2\times |V_{AdS}|$. Beyond that we encounter run-away behavior $\rho\to\infty$. \par  

However, we would like to point out that matching the KPV form of the $\overline{D3}$-induced uplift does not uniquely determine the form of the superpotential. The route determining the dependence of $W$ on $S$ described above constitutes only one possible outcome of matching KPV. The ambiguity arises from the 4D point of view by the appearance of a new mass scale due to non-perturbative K\"ahler moduli stabilization, which was absent in the reduction leading to the KPV form of the $\overline{D3}$-induced scalar potential. To see this in detail, we shall repeat the steps needed to match a candidate form of $W=W(S)$ to the KPV form of the uplift:

\begin{itemize}
		\item The classical potential for an anti-brane is determined by dimensional reduction over a classical no-scale GKP background \cite{Kachru:2002gs}.
		
		\item The SUSY-breaking anti-brane can be described within the $4D$ effective supergravity of volume stabilization in the KKLT setup by a nilpotent multiplet S which incorporates the non-linear SUSY of the anti-brane into the supergravity description. 
		
		\item The $4D$ volume stabilization generates a new scale $m_\rho^2 M_{\rm P}^2$ .
		
		\item 
		In this situation, the classical potential of the anti-brane carries over from the no-scale GKP background only in the limit $\delta V\gg m_\rho^2 M_P^2$ (i.e. comparatively weak warping) while a constant uplift potential is appropriate in the limit $\delta V\ll m_\rho^2 M_P^2$ where backreaction on the volume modulus can be neglected.
		
		\item Matching with the no-scale reduction does not uniquely fix where the nilpotent superfield $S$ appears in the superpotential $W(S,\rho)$.
		
		\item The sign of the vacuum energy as a function of increasing uplifting depends on the detailed form of $W(S,\rho)$.
\end{itemize}
The most general superpotential to leading order in the (fractional) instanton expansion is
\begin{equation}
\label{antibranesuperpot}
W=W_0+b\cdot S+A(1+c\cdot S)e^{ia\rho}\, .
\end{equation}
The special choice $c=0$, $b=e^{2\mathcal{A}_0}\mu^2$ corresponds to the standard KKLT potential. The other extreme case would be to set $b=0$. In this case the SUSY KKLT vacuum corresponding to $c=0$ cannot be uplifted to de Sitter for any value of $c$. Obviously there exist infinitely many intermediate situations.\footnote{Note that this is not the most general way to parametrize the antibrane potential (see e.g. \cite{Kallosh:2017wnt}): More generally, starting from eq. \eqref{eq:supo-KKLT} one may rescale $S\longrightarrow e^{-2\mathcal{A}_0}S$ such that the warp factor appears in the K\"ahler rather than in the superpotential as in \cite{GarciadelMoral:2017vnz}. The leading non-perturbative corrections to the anti-brane potential could then be encoded by replacing $e^{4\mathcal{A}_0}\longrightarrow e^{4\mathcal{A}_0}+\alpha(\rho,\overline{\rho})$. Only for special choices of the real function $\alpha(\rho,\overline{\rho})$ this can be transformed into a superpotential correction. We thank Susha Parameswaran for bringing this to our attention.}

Of course it would be rather surprising if the term proportional to $c$ were completely absent for the following reason. It is well known that if mobile $D3$ branes are present, the coefficient of $e^{ia\rho}$ in the superpotential is a holomorphic function of their position moduli \cite{Berg:2004ek,Baumann:2006th}. Their moduli space is therefore lifted by the same non-perturbative effects that lead to volume stabilization. If $D3$-branes modify the coefficient of the exponential term in the superpotential, one would expect an $\overline{D3}$-brane to do so as well.

However, this modification has to be quite drastic in order to have any important effect. In other words, because the coefficient $b$ is associated to the squared warp factor at the tip of the throat $e^{2\mathcal{A}_0}\ll 1$ and the coefficient $c$ is multiplied by the gaugino condensate, the effect of the term $S\, e^{ia\rho}$ will be negligible if it is further suppressed by warping. At first glance this seems unreasonable as the anti-brane sits in a strongly warped region of space-time. So, why should there exist any unwarped contributions to the superpotential? First of all, there is no $4D$ EFT reason that would forbid such a term so in the spirit of a Wilsonian effective action we should include it. If coefficients in the effective action are small, there should better exist a good reason for their smallness. We have such a reason, namely warping, only for the classical term. On the contrary, we have no a priori reason to assume smallness also of the $S$-coupling to the non-perturbative term. Still, an unsuppressed coupling of the $S$ field parametrizing a warped anti-D3-brane may sound quite counterintuitive. While we must leave a fuller discussion to later when we have introduced the full 10D description, we note here that the presence of the non-perturbative term signals the presence of a new mass scale $\sim \langle\lambda\lambda\rangle$ in the effective physics. Absent any protective symmetries or dynamical coupling suppression, such a new mass scale will communicate itself generically to all other scalar couplings, in particular the warped KK-scale. This would in turn lead one to naively expect couplings between $S$ and $\langle\lambda\lambda\rangle$ of $\mathcal{O}(1)$ following the spirit of Wilsonian EFT.

%Indeed, if an uncharged object of some tension is included at the bottom of a warped throat, one would expect its potential energy to be suppressed by four powers of the warp factor, so strictly speaking there should be no unwarped contribution to the potential. However, this misses a crucial point: The IR-warp factor itself could be corrected by the bulk non-perturbative effects. A possible change of perspective would be to interpret the combination $e^{4\mathcal{A}_{qu,0}}\equiv |b+A \,c\cdot e^{ia\rho}|^2$ as the quantum corrected IR warp factor. We do not know how to determine this correction within the framework of $4D$ EFT. However, we have no a priori reason to expect the correction parametrized by the coefficient $c$ to be suppressed by the \textit{classical} warp factor. We will come back to this issue in section \ref{10D}.

Given this, to us it appears plausible that any energy density, be it in a warped throat or in the bulk Calabi-Yau, effectively couples to the non-perturbative effects without any further warp. Hence, to us the following superpotential seems to adequately parametrize the interplay of an anti-brane with non-perturbative volume stabilization,
\begin{equation}
\label{antibranesuperpot1}
W=W_0+e^{2\mathcal{A}_{0}}\cdot \mu^2\cdot S+A(1+ c(\mu^2)\cdot S)e^{ia\rho}\, .
\end{equation}
 Again, $|\mu|^4$ is associated to the unwarped tension of the anti-brane. $e^{2\mathcal{A}_{0}}$ is the IR warp factor of the \textit{classical} ISD background while $c(\mu^2)$ is an unknown function of the anti-brane tension that we believe is \textit{not} suppressed by powers of the warp factor. Whether or not an uplift to de Sitter is possible depends on the magnitude and functional form of $c(\mu^2)$.\footnote{We consider an uplift to de Sitter successful if it does not reduce the mass-scale of the lightest modulus as compared to the pre-uplifted configuration by factors that control the perturbative expansion of string theory such as cycle volumes or the dilaton. Only with this restriction, a truncation to the lowest order K\"ahler potential necessary for moduli stabilization is consistent in presence of the uplift.}\par
The observation that the fate of the uplift depends so heavily on the details of the moduli potential can be embedded into a more systematic approach towards a de Sitter uplift: in the $4D$ effective field theory one may expand the scalar potential (before any attempt to uplift it) around its SUSY $AdS$ (or Minkowski) minimum,
\begin{equation}
V=V_0+\sum_i \frac{1}{2}m_i^2u_i^2+\cdots\, ,
\end{equation}
with squared masses $m_i^2$ (assuming canonical normalization).\par
Including an uplift means adding a further term $\delta V(u_i)$ to the potential,
\begin{equation}
V(u_i)\longrightarrow \tilde{V}(u_i)=V(u_i)+\delta V(u_i)\, .
\end{equation}
Although in general in order to determine the value of the potential at the new minimum $u^1_i$ one has to minimize the new potential $\tilde{V}$ with respect to all moduli $u_i$ one easily sees that the resulting shift in potential energy can be expanded in powers of $\delta V$ as
\begin{equation}
\label{flatteningcor.}
\tilde{V}(u_i^1)-V(u_i^0)=\delta V(u_i^0)\left(1-\sum_i\frac{1}{2}\frac{(M_{\rm P}\del_{u_i}\delta V|_{u_i^0})^2}{M_{\rm P}^2m_i^2 \delta V(u_i^0)}+\mathcal{O}((\delta V/m_i^2M_{\rm P}^2)^2)\right)\, .
\end{equation}
Evidently, for sufficiently small uplift potential $\delta V$ the back-reaction on the fields $u^i$ can be neglected and the extra potential energy simply adds up to the pre-existing one. However, in the simplest examples of moduli stabilization \cite{Kachru:2003aw} the value of the cosmological constant at the supersymmetric minimum $V_{AdS}$ is tied to the mass-scale $m_{\rho}$ of the lightest modulus
\begin{equation}
|V_{AdS}|\sim m_{\rho}^2M_{\rm P}^2\, .
\end{equation}
Clearly, in order to uplift such vacua to de Sitter, an uplift potential $\delta V$ has to be added that is at least of the order of $|V_{AdS}|$. In this case the expansion in \eqref{flatteningcor.} is not under parametric control and the strength of back-reaction effects heavily depends on the detailed moduli dependence of $\delta V$. For the uplifted KKLT potential a strong back-reaction effect does not occur due to the proposed power-law dependence of $\delta V\equiv |\mu^4|/12 \text{Im}(\rho)^2$. In this case the first correction in \eqref{flatteningcor.} is suppressed by a volume factor $(a\,\text{Im}(\rho))^{-2}\ll 1$. In contrast, for an exponential dependence $\delta V \sim e^{-2a\,\text{Im}(\rho)}$ corrections in \eqref{flatteningcor.} are unsuppressed. This is precisely what a sufficiently dominant term $\propto c(\mu^2)$ in \eqref{antibranesuperpot1} would entail.\par
We conclude that if the vacuum potential of the SUSY $AdS_4$ vacua can be parametrically decoupled from the mass-scale of the lightest modulus,
\begin{equation}
|V_{AdS}|\ll m_{\rho}^2 M_{\rm P}^2\, ,
\end{equation}
the required uplift can be small enough such that backreaction effects are negligible and thus, de Sitter uplifts are generically possible. Conversely, if this is not the case detailed knowledge about the moduli dependence of $\delta V(u_i)$ is required in order to determine the uplift in potential energy beyond parametric estimates.\par

Finally, we summarize the upshot of this section and set the goal for the rest of the paper:
\begin{itemize}
	\item In order to determine if an anti-brane can uplift the SUSY KKLT vacua to de Sitter we need more information about the interplay of the anti-brane with non-perturbative effects. 
	
	\item In the remainder of the paper we will determine only the sign of the cosmological constant in the presence of an anti-brane uplift using a $10D$ description. This will allow us to constrain the coefficients of $W(S,\rho)$ only to the extend that the sign of the $4D$ vacuum energy as determined from $W(S,\rho)$ must be matched to the $10D$ result. In particular we will \textit{not} determine the form of the scalar potential $V(\rho)$ away from its minimum. This important task that would \textit{uniquely} fix the coefficients in the superpotential we leave for future work.
\end{itemize}
%\newpage

\section{Some higher dimensional considerations}

In view of the surprising difficulties that one usually encounters when trying to construct consistent de Sitter vacua in string theory we find it worthwhile to investigate if the seemingly conspirative modification of the $4D$ effective field theory that would prevent an uplift to de Sitter indeed occurs.\par 
Clearly the cleanest way to do this would be to derive the correct effective field theory of the volume modulus together with all its SUSY breaking states from first principles. Due to the obvious difficulty of this approach we will opt for another one. Instead of deriving the off-shell $4D$ effective potential we will constrain it using a $10D$ on-shell description of the non-perturbative effects as well as the uplift. Before turning to the $10D$ setup we  study some general aspects of compactifications that will later be relevant and use them to explain this notion of an \textit{on-shell} description. This concept will be clarified in terms of a toy model that will also be useful to develop a physical intuition that is applicable to the case of type IIB string theory. 

\subsection{On-shell vs off-shell potentials}
\label{OnOff}

Let us first explain what we mean by on-shell and off-shell potentials. We refer to the usual $4D$ potential that is determined by straightforward dimensional reduction as the \textit{off-shell potential}. It is easily determined from the higher dimensional Einstein equations as follows: for a theory of $D=d+4$ dimensional gravity and further fields compactified to $4D$ de Sitter, Minkowski or anti de Sitter, the most general $D$-dimensional metric ansatz is
\begin{equation}
ds^2=e^{2\mathcal{A}(y)}\tilde{g}^4_{\mu\nu}(x)dx^{\mu}dx^{nu}+g^{d}_{mn}(y)dy^mdy^n\, ,
\end{equation}
with warp factor $e^{\mathcal{A}}$, $4D$ coordinates $x^{\mu}$ and internal coordinates $y^m$. The higher dimensional Einstein equations
\begin{equation}
R_{MN}-\frac{1}{2}g_{MN}R=T_{MN}\, 
\end{equation}
determine the effective $4D$ potential in terms of the internal curvature $R_d$ and the higher dimensional energy momentum tensor $T_{MN}$,
\begin{equation}
\label{4dpotential_offshell}
V\cdot M_{\rm P}^{-4} =  \dfrac{1}{4 } \tilde{R}_{4} \cdot M_P^{-2} =\mathcal{V}_w^{-2}M^d\int d^d y\sqrt{g^d}e^{4\mathcal{A}}\frac{(- \frac{1}{4}T^{\mu}_{\mu}-\frac{1}{2}R_d)}{M^2}\, ,
\end{equation}
where $M$ is the D-dimensional Planck mass, $\mathcal{V}_w=M^d\int d^d y\sqrt{g^d}e^{2\mathcal{A}}$ is the warped volume and $M_{\rm P}^2=M^2\mathcal{V}_w$ is the $4D$ Planck mass. $4D$ vacua of the higher-dimensional theory correspond to local minima of this potential which encodes the value of the cosmological constant as well as the scalar mass-spectrum.\par  
In many instances it is hard to determine this potential in full explicitness and it may be very useful to use an alternative expression for the $4D$ potential that is only valid when all geometric moduli are assumed to be \textit{at} their minimum. One may obtain such an expression by considering the trace-reversed Einstein equations that relate the internal space Ricci scalar $R_d$ to the energy momentum tensor. The result is
\begin{equation}
\label{4dpotential_onshell}
V\cdot M_{\rm P}^{-4}=\mathcal{V}_w^{-2}M^d\int d^d y \sqrt{g^6}e^{4\mathcal{A}}\frac{(D-6)T^{\mu}_{\mu}-4T^{m}_m}{4(D-2)M^2}\, .
\end{equation}
It is important to realize that in writing this expression one has integrated out all geometric moduli, that is the potential is only valid \textit{at the minimum} of the off-shell potential \eqref{4dpotential_offshell} where the two coincide \cite{Giddings:2005ff}.  This has several consequences: for example,  if a higher-dimensional source contributes an energy momentum tensor $\delta T_{MN}$ with
\begin{equation}
(D-6)\delta T^{\mu}_{\mu}-4\delta T^m_m< 0\, ,
\end{equation}
one must not conclude that the $4D$ cosmological constant decreases when the source is included since all other terms in the full energy momentum tensor $T_{MN}$ will be perturbed even \textit{to leading order} in $\delta T_{MN}$. Although this means that any on-shell potential has to be treated carefully one can derive powerful statements from it. For example, in the absence of \textit{any} source  satisfying
\begin{equation}
\label{necessary_cond_dS}
(D-6)T^{\mu}_{\mu}-4T^m_m\geq 0\, ,
\end{equation}
the $4D$ vacuum energy can never be positive. Using this, one may show that if one only allows for $p$-form fluxes with $1\leq p \leq D-1$ and localized objects of positive tension and co-dimension $\geq 2$, de Sitter solutions are ruled out \cite{Maldacena:2000mw}. Further details are given in Appendix \ref{App:Contr.to.onshell.pot}.  \par 
This means that in many interesting examples the on-shell potential encodes subtle back-reaction effects that correspond to the correction terms in \eqref{flatteningcor.}: whenever a source of positive higher-dimensional energy is turned on that does not satisfy the condition \eqref{necessary_cond_dS} in a compactification that is stabilized by sources that also do not satisfy the condition, higher order corrections in \eqref{flatteningcor.} conspire to keep the overall potential energy negative. 

\subsection{A simple example: Freund-Rubin compactification}
We will now demonstrate this behavior using the well known Freund-Rubin compactification. This is a  $6D$ theory compactified  on a $ S^2$, with the $S^2$ stabilized by $2$-form fluxes \cite{Freund:1980xh}. The $6D$ action is
\begin{equation}
S_6=\frac{M^4}{2}\int\left(*R_6-\frac{1}{2}F_2\wedge*F_2\right)\, ,
\end{equation}
with a $2$-form field-strength $F_2=dA_1$. The equations of motion/Bianchi identity are
\begin{equation}
R_{MN}=\frac{1}{2}F_{MP}{F_N}^P-\frac{1}{8}g_{MN}|F_2|^2\, ,\quad dF_2=0=d*F_2\, .
\end{equation}
These admit a solution where the $6D$ geometry is a product $AdS_4\times S^2$ and the $S^2$ is threaded by $N$ units of $2$-form flux $F_2=\frac{N}{2q}\omega_2$. Here, $\omega_2$ is the volume form of the $S^2$ normalized to $\displaystyle\int_{S^2} \omega_2=4\pi$ and $q$ is the  $U(1)$ charge of the particle that couples electrically to $A_1$ with smallest charge. The $S^2$ radius is fixed at
\begin{equation}
\label{S2radius}
L_0^2=\frac{3N^2}{32q^2}\, ,
\end{equation}
and in agreement with \eqref{4dpotential_onshell} the $4D$ vacuum energy reads 
\begin{equation}
\label{S^2comp.pot}
V\cdot M_{\rm P}^{-4}=\mathcal{V}^{-2}M^2\int d^2 y \sqrt{g^{S^2}}\frac{(-2|F_2|^2)}{16M^2}=-(12\pi M^4L_0^4)^{-1}\, .
\end{equation}
One may try to uplift the four dimensional vacuum energy by adding a number $N_3$ of three-branes of positive tension $T_3$ smeared over the internal two-sphere. Clearly they are a source of  energy density and therefore the expectation is that they give rise to an increase of the 4 dimensional vacuum energy. However,   their energy momentum tensor satisfies $(D-6)T^{\mu}_{\mu}-4T^{m}_m=0$  and so there is no new contribution under the integral in  \eqref{4dpotential_offshell}. Even without knowing the full solution to the $6D$ equations of motion we can hence express the $4D$ vacuum energy as a function of the a priori unknown size of the two-sphere:
\begin{equation}
\label{on-shell-6d}
V\cdot M_{\rm P}^{-4}=\mathcal{V}^{-2}M^2\int d^2 y \sqrt{g^{S^2}}\frac{(-2|F_2|^2)}{16M^2}=-\frac{M^2L_0^2}{12\pi M^6 L_1^6}\, ,
\end{equation}
where $L_0$ is given in eq. \eqref{S2radius} and $L_1$ is the adjusted length-scale of the two-sphere. Here, the wonders of on-shell potentials manifest themselves explicitly for the first time: The source of uplift in potential energy, the three-brane tension, has apparently disappeared from the integrand of the on-shell potential. It does however appear \textit{implicitly} through the dependence of the $S^2$-radius on the three-brane tension. This way of parameterizing the $4D$ vacuum energy in terms of a single unknown quantity (in this case the radius) should be kept in mind as it will play an important role in the remainder of the paper. From eq. \eqref{on-shell-6d} one may already follow that no matter how much three-brane tension is added, the vacuum energy cannot increase beyond zero but at most approaches zero from below. In this simple $6D$ case one can do better and from the internal Einstein equations determine $L_1$ as a function of the three-brane tension:
\begin{equation}
L_1^2=\left(1-\mathcal{T}_3\right)^{-1}L_0^2\, , \quad \text{with}\quad \mathcal{T}_3\equiv \frac{N_3T_3}{4\pi M^4}\, .
\end{equation}
Plugging this into eq. \eqref{on-shell-6d} we see that the vacuum energy adjusts and as expected to leading order in the dimensionless tension $\mathcal{T}_3$ the change in potential energy corresponds precisely to the three-brane tension
\begin{equation}
\delta(V\cdot M_{\rm P}^{-4})=\frac{\mathcal{T}_3}{4\pi L_0^4M^4}+\mathcal{O}(\mathcal{T}_3^2)=\frac{N_3 T_3}{M_{\rm P}^4}+\mathcal{O}(\mathcal{T}_3^2)\, .
\end{equation}
Of course, as we increase the three-brane tension such that $\mathcal{T}_3\longrightarrow 1$ the $4D$ vacuum energy increases, but  the higher order terms   become relevant and will prevent an uplift to de Sitter. Therefore,  the 4D vacuum energy will approach  zero from below,   the limit being $\mathcal{T}_3=1$, where  the   $S^2$ decompactifies. Note that as predicted by \eqref{flatteningcor.} the expansion parameter that controls back-reaction is given by $\delta V/m_{KK}^2M_{\rm P}^2$ because the KK-scale is the mass of the lightest modulus.\par
One might be concerned that three-branes and two-form fluxes share an intrinsic property making them unsuitable uplifting ingredients because they appear with the wrong sign under the integral of \eqref{4dpotential_onshell}. Of course, if one \textit{only} includes these ingredients in the compactification de Sitter solutions are ruled out \cite{Maldacena:2000mw}. But as we now demonstrate it is enough to include also a positive $6D$ c.c., or equivalently a five-brane of tension $T_5\equiv \mathcal{T}_5 M^6$, for an uplift to de Sitter \textit{by three-branes or fluxes} to be possible (see also \cite{Saltman:2004jh} for related conclusions). In this case the size of the $2$-sphere is bounded from above via
\begin{equation}
l^2\equiv L^2M^2=\frac{1-\mathcal{T}_3}{\mathcal{T}_5}\left(1-\sqrt{1-\frac{3}{16}\frac{\mathcal{T}_5n^2}{(1-\mathcal{T}_3)^2}}\right)\leq \frac{1-\mathcal{T}_3}{\mathcal{T}_5}\, ,
\end{equation}
where $n\equiv N\cdot M/q$ corresponds to the number of two-form flux quanta. When $n^2>n_{max}^2\equiv  \frac{16}{3}\mathcal{T}_5^{-1}(1-\mathcal{T}_3)^2$ the sphere decompactifies.
Thus, in order for the curvature of the sphere to be sub-planckian we need both a small positive $6D$ cosmological constant $\mathcal{T}_5\ll 1$ as well as a large number of two-form fluxes $n$.\par
The on-shell potential according to \eqref{4dpotential_onshell} reads
\begin{equation}
V\cdot M_{\rm P}^{-4}=\frac{1}{16\pi}\left[\frac{2\mathcal{T}_5}{l^2}-\frac{n^2}{8l^6}\right]\, .
\end{equation}

\begin{figure}[t!]
\begin{center}
	\includegraphics[keepaspectratio, width=8cm]{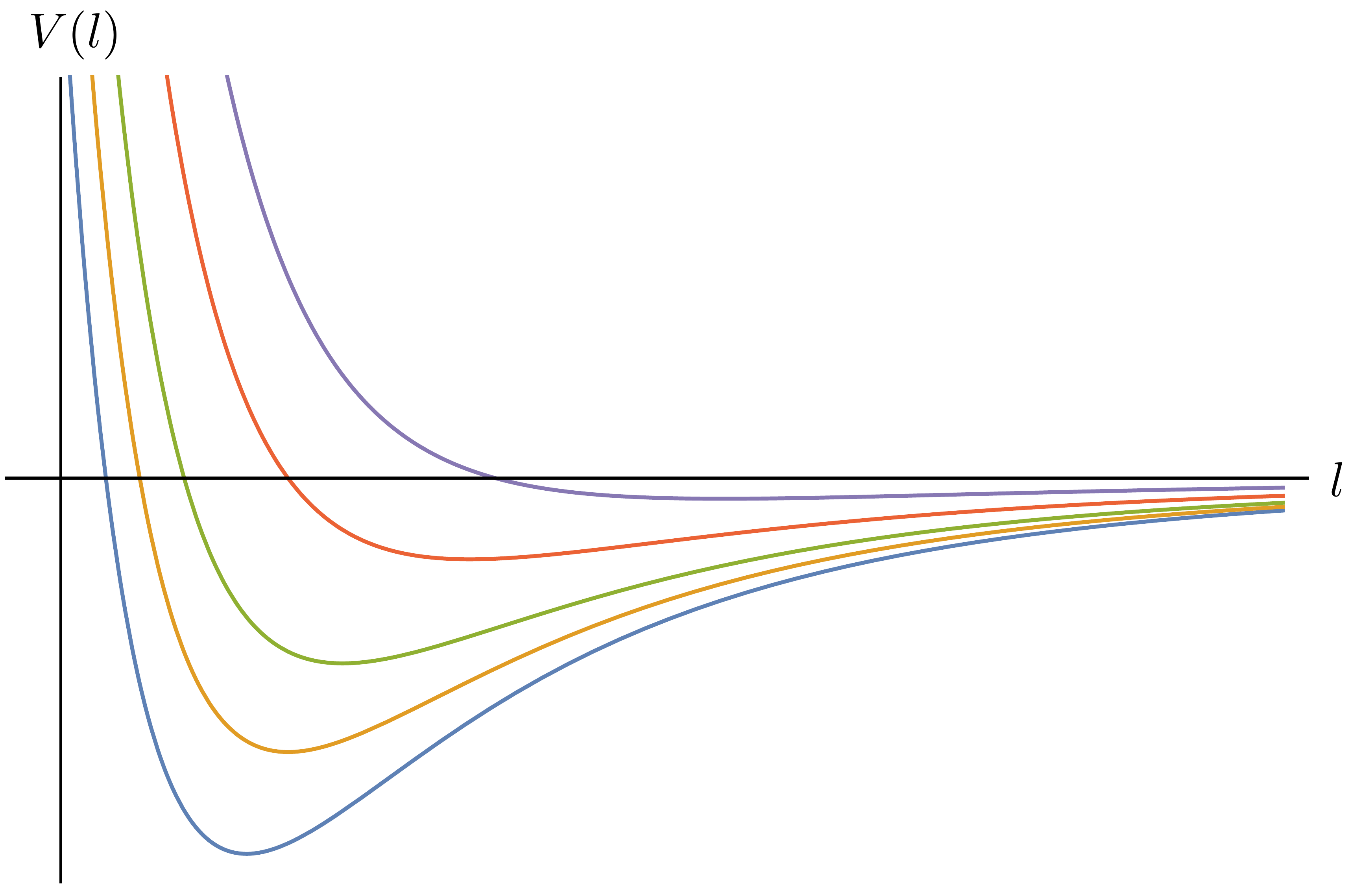}
	\includegraphics[keepaspectratio, width=8cm]{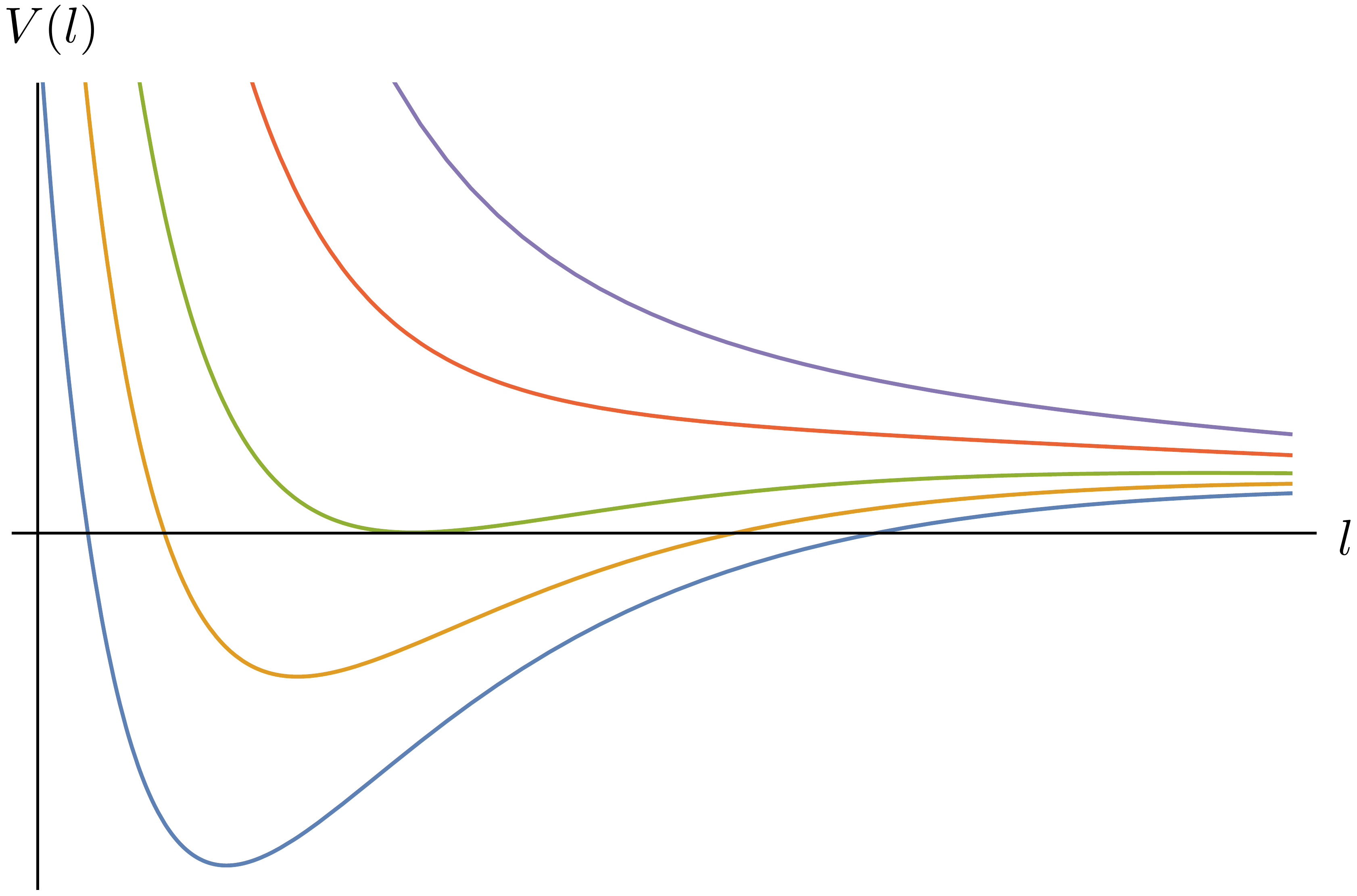}
\end{center}
	\caption{Left: Off-shell potential $V(l)$ for the $S^2$ volume modulus in the case without a $6D$ c.c. for $n=25$ flux units and different values of the dimensionless three-brane tension $\mathcal{T}_3$:  $\mathcal{T}_3=0$ in blue, $\mathcal{T}_3=0.1$ in yellow, $\mathcal{T}_3=0.21$ in green, $\mathcal{T}_3=0.4$ in red and $\mathcal{T}_3=0.6$ in purple. As can be seen, the more energy density, the higher the vacuum energy, but the flattening prevents the minimum to go above zero. Right: Off-shell potential $V(l)$ for the $S^2$ volume modulus in the case with a $6D$ c.c. for $n=25$ flux units, $\mathcal{T}_5=0.004$ and different values of the dimensionless three-brane tension $\mathcal{T}_3$:  $\mathcal{T}_3=0$ in blue, $\mathcal{T}_3=0.1$ in yellow, $\mathcal{T}_3=0.21$ in green, $\mathcal{T}_3=0.4$ in red and $\mathcal{T}_3=0.6$ in purple. This time  it is also possible to find de Sitter minima once enough three-brane tension has been added. }
	\label{fig:toy.model}
\end{figure}

Clearly the no-go theorem of Maldacena and Nunez~\cite{Maldacena:2000mw} is evaded and de Sitter vacua are possible although not generic.\footnote{As an aside we note that the de Sitter vacua with largest possible cosmological constant lie along the line of marginal stability $n=n_{max}$ for which $l^2=l_{max}^2\equiv \frac{1-\mathcal{T}_3}{\mathcal{T}_5}$. The value of the c.c. is given by
	\begin{equation}
	V_{max}\cdot M_{\rm P}^{-4}=\frac{\mathcal{T}_5}{12\pi}\underbrace{\frac{\mathcal{T}_5}{1-\mathcal{T}_3}}_{=l_{max}^{-2} \ll 1}\lesssim l_{max}^{-4}\, .
	\end{equation}
	It is of order the higher dimensional c.c. times the usual volume suppression factor. Note however that the required tuning of the $6D$ c.c. amounts to a further volume suppression $V_{max}\lesssim l^{-4}$.} We give a concrete example in Figure \ref{fig:toy.model}. \par 
For completeness we also give the off-shell potential (plotted in Figure \ref{fig:toy.model})
\begin{equation}
V\cdot M_{\rm P}^{-4}(l)=\frac{1}{16\pi}\left[\frac{4\mathcal{T}_5}{l^2} +\frac{n^2}{4l^6}-\frac{4(1-\mathcal{T}_3)}{l^4}\right]\, .
\end{equation}
Evidently, the flattening behavior observed for the case with only fluxes and three-branes does not exhibit any intrinsic feature of branes and fluxes but is merely a property of the simple scheme of moduli stabilization. By including a positive $6D$ c.c. it was possible to decouple the lightest modulus mass from the value of the $4D$ c.c. such that a small perturbation could uplift it to $4D$ de Sitter.
%%%%%%%%%%%%%%%%%%%%%%%%%%%%%%%%%%%%%%%%%%%%%%%%%%%%%%%%%%%%%%%%%%%%%%%%%%

\section{The $10D$ Perspective}\label{10D}
We would now like to study moduli stabilization and the uplift to de Sitter space from a ten-dimensional point of view. The classical part is well understood: the Gukov-Vafa-Witten superpotential can be lifted to the ten-dimensional three-form potential of type IIB supergravity \cite{Gukov:1999ya} and the $4D$ SUSY conditions that determine the three-form fluxes to be of Hodge-type $(2,1)$ lift to the $10D$ SUSY conditions of \textit{$B$-type} \cite{Grana:2004bg}. Furthermore the $4D$ scalar potential is minimized precisely when the $10D$ equations of motion are solved by the imaginary self-dual (ISD) solutions of \cite{Giddings:2001yu}.\par 
An analogous $10D \longleftrightarrow 4D$ correspondence of K\"ahler moduli stabilization is somewhat harder to establish, both conceptually as well as technically:\footnote{We thank Arthur Hebecker for a very helpful discussion concerning this point.}  the dynamical origin of the exponential superpotential is the condensation of gaugino bilinears in the $4D$ SYM gauge theory (or euclidean $D3$ brane instantons). The scale below which the condensation occurs is the dynamical scale of the gauge-theory which typically lies far below the Kaluza-Klein scale. So, how can it be possible even in principle to include the non-perturbative effects in a higher-dimensional setup? First, there certainly exist geometrical setups compatible with the correct order of scales: an example is  that of an `anisotropic' Calabi-Yau space in which the four-cycle that the $7$-branes wrap is much smaller than the typical length-scale of the transverse space \cite{Baumann:2010sx}. In this case the non-perturbative scale of gaugino condensation can lie far below the Kaluza-Klein scale of the four-cycle and at the same scale as the transverse Kaluza-Klein scale. Another  situation of this type corresponds to a compactification space that is equipped with warped throats of significant warping. In this case the warped Kaluza-Klein scale lies exponentially below the bulk KK-scale.\par 
There has however been crucial progress in recent years in establishing a far more general ten dimensional picture of gaugino condensation \cite{Baumann:2006th,Baumann:2007ah,Dymarsky:2010mf,Heidenreich:2010ad}. First, note that if a mobile $D3$-brane is present, the classical moduli space of the world-volume scalars is identified with the compactification geometry. In the absence of non-perturbative effects there is no potential for the world-volume scalars and the internal geometry can thus be probed at arbitrarily small energies. Thus, even if non-perturbative effects generate a potential for the world volume scalars one may probe the (quantum-deformed) internal geometry at scales that lie far below the KK-scale. With this in mind one should be able to effectively describe the SUSY vacua with non-perturbative K\"ahler stabilization by the $10D$ equations of motion, corrected at order of the value of the gaugino condensate $\langle \lambda \lambda \rangle$.\par 
So, how is the $10D$ supergravity corrected in order to account for the non-perturbative effects? Remarkably, the following simple prescription advocated by the authors of \cite{Frey:2005zz,Dymarsky:2010mf,Baumann:2010sx} seems to capture at least the leading order effects:
\begin{enumerate}
	\item Use the classical type IIB supergravity together with the DBI and CS actions for localized objects to quadratic order in the worldvolume fermions.
	\item Solve the $10D$ equations of motion, assuming that the fermion bilinear that corresponds to the $7$-brane gaugino is non-vanishing.
\end{enumerate}
Clearly this approach needs to be justified. For this let us consider the non-perturbative lifting of the $D3$-brane position moduli space. 
This can be studied from different angles. The first is the standard $4D$ perspective. In compactifications with both $D7$-branes and mobile $D3$ branes the gauge-kinetic function of the $D7$ brane gauge theory depends on the open-string $D3$-brane position moduli $z^i$ via one-loop open string threshold corrections which were calculated explicitly for a $T^4/\mathbb{Z}_2\times T^2$ orientifold of type IIB string theory \cite{Berg:2004ek}. Then, at low energies the non-perturbative superpotential is a function of the position moduli $z^i$ which obtain a potential. The open string calculation of \cite{Berg:2004ek} was perfectly matched with a dual \textit{closed string} calculation in \cite{Baumann:2006th,Baumann:2007ah} as follows: a mobile $D3$ brane treated as a classical localized source in the $10D$ supergravity induces corrections to the volume of the $4$-cycle that the $D7$-branes wrap which determines the gauge-coupling of the $D7$ gauge theory.  Again, the $D3$ position moduli enter the non-perturbative superpotential in the $4D$ EFT\footnote{See also \cite{Koerber:2007xk,Koerber:2008sx} for a derivation using the language of \textit{generalized complex geometry}.}. The closed string computation is particularly useful as it readily generalizes beyond simple toroidal orientifolds. In particular, for a stack of $N$ $D7$-branes with holomorphic embedding equation $h(z)=0$, the gauge-kinetic function $f(\rho,z^i)$ of the $D7$ gauge theory depends on the volume modulus $\rho$ as well as the $D3$ position moduli $z^i$  \cite{Baumann:2006th}
\begin{equation}
\label{gauge_kin_function}
f(\rho,z)=\rho+\frac{\ln h(z)}{2\pi i}\, .
\end{equation}
Using this dependence of the gauge-kinetic function $f$ on the $D3$-brane position moduli one may determine the $4D$ non-perturbative superpotential to be
\begin{equation}
\label{nonpert_superpot}
W\propto e^{\frac{2\pi i}{N}f}=h(z)^{1/N}e^{\frac{2\pi i}{N}\rho}\, .
\end{equation}
So far, classical $10D$ physics has been used only to obtain the gauge kinetic function \eqref{gauge_kin_function} while the generation of a non-trivial potential for the $D3$-brane moduli is determined entirely within $4D$ effective field theory. Crucially these two steps could be separated because the classical back-reaction of a $D3$-brane on the classical $10D$ supergravity solution is finite. This is clearly not the case for an $\overline{D3}$-brane, so the quantum corrected $10D$ action is needed. So how can it be deduced? The key points were derived in  \cite{Baumann:2010sx}, where the authors analyzed the generation of a non-trivial \textit{classical} potential for the position moduli of $D3$ branes in ISD backgrounds subject to harmonic non-ISD perturbations. Crucially, it was shown that in conifold backgrounds every superpotential that can be written down for the position moduli in the $4D$ effective field theory can be matched to a non-compact \textit{classical} $10D$ supergravity solution such that the scalar potentials coincide. Hence, the quantum corrected $10D$ supergravity that reproduces the correct $D3$ brane potential is only corrected by terms that are localized away from the warped throat. Such localized terms are necessary because the entirely uncorrected type IIB supergravity equations do not admit static non-ISD perturbations in the compact case due to the \textit{global} constraints of  \cite{Giddings:2001yu}. \par
\begin{figure}[t!]
	\centering
	\includegraphics[keepaspectratio, width=16cm]{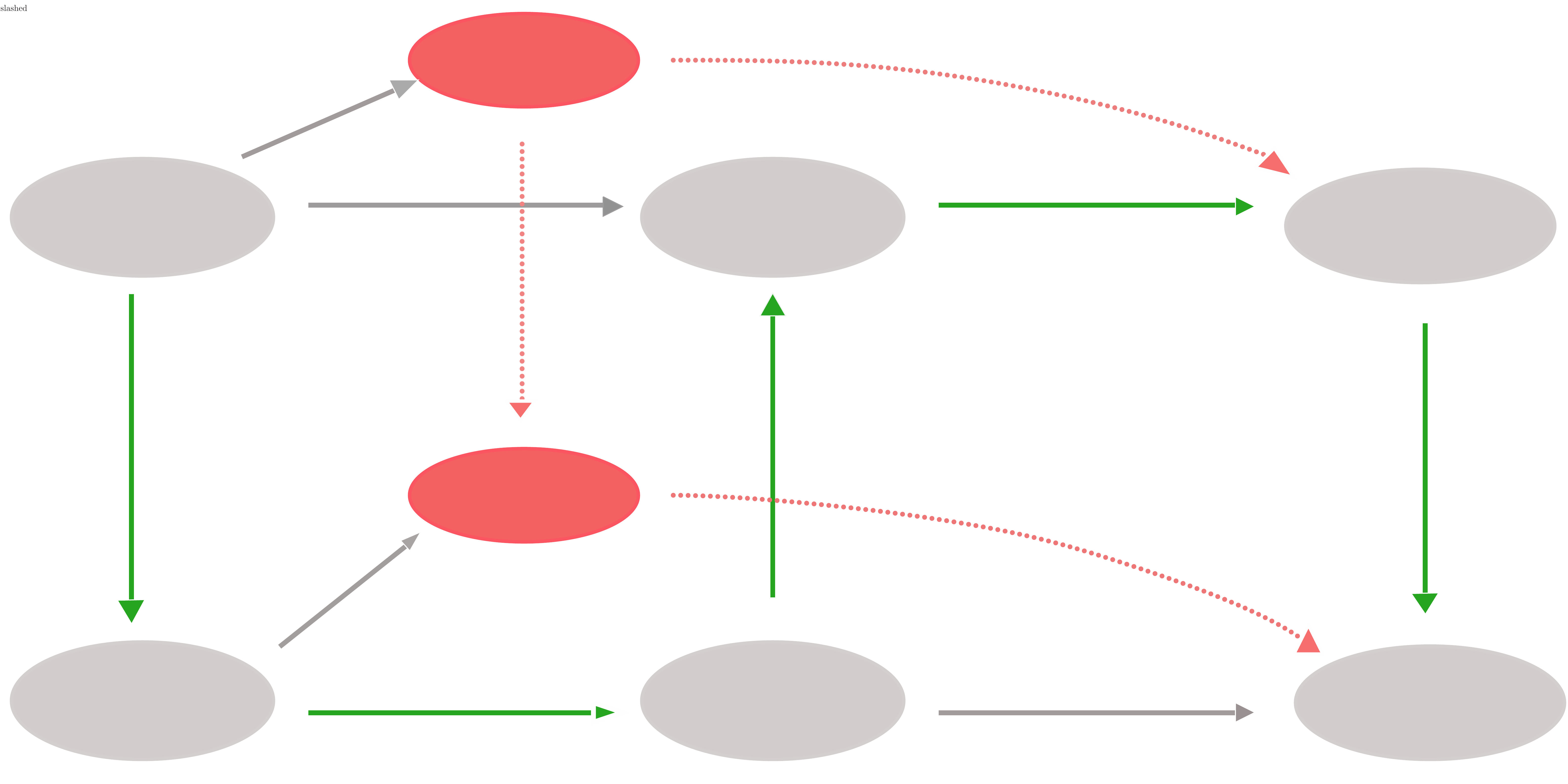}
	\caption{From the classical $10D$ supergravity to the quantum $4D$ theory with broken supersymmetry. Green arrows denote steps that can be followed through unambiguously and with reasonable amount of control. The dimensional reduction of the classical $10D$ theory is well understood. Incorporating non-perturbative quantum effects in the $4D$ EFT can be performed with reasonable amount of control as well. The de Sitter uplift of the supersymmetric KKLT vacua within $4D$ EFT suffers from ambiguities as explained in section \ref{KKLT:uplift}. For this reason we follow the authors of \cite{Baumann:2006th,Baumann:2007ah,Dymarsky:2010mf,Baumann:2010sx} in first lifting the SUSY KKLT vacua to $10D$. Since at this stage we have a quantum deformed $10D$ action we may include the SUSY breaking effects of an anti-brane and thereafter dimensionally reduce the $10D$ potential to $4D$. This prescription avoids the intermediate classical $10D$/$4D$ descriptions with a SUSY breaking runaway potential (marked in red).}\label{fig:From10dto4d}
\end{figure}

 It is tempting to identify these localized terms with the terms in the $7$-brane action that are proportional to the gaugino bilinear $\langle \lambda\lambda \rangle$.
Indeed, the superpotential \eqref{nonpert_superpot} can be encoded in so-called \textit{series I} three-form flux
\begin{equation}
(G_3)_{i\bar{j}\bar{k}}\propto \langle \lambda\lambda\rangle \nabla_i \nabla_l \text{Re}(\ln h(z) )g^{l\bar{m}}\bar{\Omega}_{\bar{m}\bar{j}\bar{k}}\, ,
\end{equation}
where $\Omega$ is the holomorphic three-form of the Calabi-Yau \cite{Baumann:2010sx}. This is precisely the perturbation of three-form fluxes that is sourced by the fermionic bilinear term in the $D7$-brane action (extrapolated to the non-abelian case). Guided by this  non-trivial consistency check we believe that the relevant details of non-perturbative volume stabilization are indeed captured by the classical $10D$ supergravity action assuming a non-vanishing expectation value of the gaugino bilinear. While (by construction) the $10D$ and $4D$ pictures can be used equivalently to determine the non-perturbative $D3$-brane position moduli potential (i.e. the back-reaction of the $D3$-brane on the $4$-cycle size), the $10D$ approach allows us to also incorporate the back-reaction of an $\overline{D3}$-brane unambiguously (for a diagrammatic representation of the
argument, see Figure \ref{fig:From10dto4d}).\par
Based on this conclusion, we will investigate whether or not the non-perturbative AdS vacua of KKLT can be uplifted to de Sitter.
The approach is to compactify the (quantum corrected) $10D$ theory and identify the terms in the corresponding on-shell potential to find a $10D$ picture of KKLT moduli stabilization, followed by the (partial) uplift.
%%%%%%%%%%%%%%%%%%%%%%%%%%%%%%%%%%%%%%%%%%%%%%%%%%%%%%%%%%%%%%%%%%%%%%%%%%
\subsection{Stabilization of Complex Structure Moduli: ISD solutions}\label{10dcsstab}
Flux compactifications of type IIB string theory were pioneered by Giddings, Kachru and Polchinski (GKP) \cite{Giddings:2001yu}. They showed that in the limit of dilute three-form fluxes the Gukov-Vafa-Witten superpotential descends from the $10D$ potential of type IIB string theory by straightforward dimensional reduction. Moreover, even for non-dilute fluxes, the $4D$ solutions obtained from minimizing the GVW superpotential lift to full $10D$ solutions, the imaginary self-dual (ISD) solutions. These feature exponential warping thus realizing the proposal of Randall and Sundrum \cite{Randall:1999ee} to naturally generate large hierarchies. \par 
Following their discussion one may derive an on-shell expression for the $4D$ potential that will prove extremely useful (see appendix \ref{App:TypeIIB} for details). Starting from the 10D Einstein frame action of Type IIB supergravity, we make a warped compactification ansatz for the metric
\begin{equation}
ds^2=e^{2\mathcal{A}(y)}\tilde{g}^{(4)}_{\mu\nu}(x)dx^{\mu}dx^{\nu}+ g^6_{mn}(y)dy^mdy^n\, ,
\end{equation}
and write the self dual 5-form flux as $ F_5=(1+\star_{10})d\alpha (y)\wedge \sqrt{-\tilde{g}^{(4)}}d^4x$.
The corresponding trace-reversed Einstein equations allow to write the 4D Ricci scalar $\tilde{R}_4$ in terms of the matter content on the compactification. The result can be combined with the Bianchi identity for $F_5$ such that \cite{deAlwis:2003sn}
\begin{equation}
\label{GKPnogo-maintext}
\begin{split}
\tilde{\nabla}^2 \Phi^{-}=&\tilde{R}_{4D}+\frac{e^{2\mathcal{A}}}{\text{Im}(\tau)}|G_3^-|^2+e^{-6\mathcal{A}}|\del \Phi^-|^2+ e^{2\mathcal{A}}\frac{\Delta^{loc}}{2\pi}\, ,
\end{split}
\end{equation}
where
\begin{equation}
G_3^{\pm}\equiv\frac{1}{2}(*_6\pm i)G_3\, ,\quad \Phi^{\pm}\equiv e^{4\mathcal{A}}\pm\alpha \, , \quad \text{and}\quad \Delta^{loc}\equiv\frac{1}{4}\left(T^m_m-T^{\mu}_{\mu}\right)^{\text{loc}}- T_3\rho_3^{\text{loc}}\, .
\end{equation}
Here  $G_3$ is the complexified three-form $F_3-\tau H_3$. Moreover $T_{MN}^{loc}$ and $T_3\rho_3^{loc}$ are the energy momentum tensor and $D3$-brane charge density of localized objects. We work in units such that $(M_{{\rm P},10D})^8=4\pi$. \par
By assuming the existence of a solution and demanding its global consistency one obtains an expression for the on-shell potential by integrating the above equation over the internal space
\begin{equation}
\label{4dpotentialIIB}
V\cdot M_{\rm P}^{-4}=\int \frac{d^6 y\sqrt{g^6}}{16\pi\mathcal{V}_w\tilde{\mathcal{V}}_{w}} \left[-\,e^{8\mathcal{A}} \frac{\Delta}{2\pi}-|\del \Phi^{-}|^2\right]\, ,
\end{equation}
 where 
\begin{equation}
\begin{split}
\Delta&\equiv 2\pi\frac{|G_3^-|^2}{\text{Im}(\tau)}+\Delta^{loc}\, ,\quad  \tilde{\mathcal{V}}_{w}\equiv\int_{M_6}d^6y\sqrt{g^6}\, e^{6\mathcal{A}}\, ,\quad  \text{and}\quad \mathcal{V}_w\equiv\int_{M_6}d^6y\sqrt{g^6}\, e^{2\mathcal{A}}\, .
\end{split}
\end{equation} 
From the on-shell potential \eqref{4dpotentialIIB} it follows immediately that as long as all localized sources satisfy $\Delta^{loc}\geq 0$ the unique classical Minkowski solutions of type IIB string theory are the ISD solutions,
\begin{equation}
G_3^-=\Delta^{loc}=\tilde{R}_{4D}=\Phi^{-}=0 \, .
\end{equation}
Under the same assumption de Sitter solutions are ruled out as well. Therefore, a necessary condition for realizing $4D$ de Sitter solutions is that there exists at least one localized object that satisfies $\Delta^{loc}<0$.\par 
It is important to note that the ISD solutions also match the $4D$ no-scale behavior: the volume modulus remains unfixed and corresponds to an overall rescaling of regions of weak warping, leaving strongly warped regions approximately invariant \cite{Giddings:2005ff}. Therefore, the inclusion of any further sources of positive potential energy cannot lead to a stable solution but must rather lead to decompactification. Therefore, in order to discuss the inclusion of further sources of positive energy, one first needs to incorporate K\"ahler moduli stabilization.

%%%%%%%%%%%%%%%%%%%%%%%%%%%%%%%%%%%%%%%%%%%%%%%%%%%%%%%%%%%%%%%%%%%%%%%%%%
\subsection{KKLT K\"ahler moduli stabilization: The $10D$ perspective}\label{10dKstab}
%%%%%%%%%%%%%%%%%%%%%%%%%%%%%%%%%%%%%%%%%%%%%%%%%%%%%%%%%%%%%%%%%%%%%%%%%%

As explained in the introduction of this section we will determine the effective on-shell potential \eqref{4dpotentialIIB} by dimensionally reducing the $10D$ action of type IIB supergravity in the presence of a non-vanishing value of the gaugino condensate. In order to do so the following quantities have to be evaluated,
\begin{enumerate}
	\item The value of $\Delta^{loc}$ induced by the non-vanishing gaugino bilinear that appears in the $D7$ brane action.
	\item The back-reaction on the three-form fluxes $G_3$ and $\Phi^-$.
\end{enumerate}
We evaluate these quantities only in the bulk Calabi-Yau where the effects of warping and three-form fluxes are volume suppressed. We will neglect these effects and thus work to leading order in an inverse volume expansion. Moreover we assume the $7$-brane configuration to be in the Sen-limit \cite{Sen:1996vd}, i.e. $4$ $D7$-branes on top of an $O7$-plane. In this case there is no classical $10D$ back-reaction on the Calabi-Yau geometry and the axio-dilaton is constant. The gauge group is $SO(8)$.\par
The calculation is a somewhat tedious but straightforward exercise that has been partially done by the authors of \cite{Baumann:2010sx,Dymarsky:2010mf}. We have provided the detailed derivation in  appendix \ref{App:GauginoCondensateBackr.and.contr.to.onshell.cc} and merely quote the result here: the piece in the action responsible for the perturbation of the ISD background is\footnote{Note that we extrapolate the abelian $D7$-brane term to the non-abelian case. $\lambda\lambda$ will be shorthand for $\Tr{\lambda^{\alpha}\lambda_{\alpha} }$. We would like to stress that the most powerful argument for the validity of this approach comes not from the fact that this term can be obtained from the (non-abelian) seven-brane action but rather from the $10D\longleftrightarrow 4D$ matching of the $D3$-brane position moduli potential. Therefore the extrapolation of the non-abelian $7$-brane action is no further reason of concern for us here.}
\begin{equation}
\label{llD7action}
S_{D7}\supseteq\int_{M_{10}} \pi \delta^{(0)}_De^{\phi/2}e^{-4\mathcal{A}}\frac{\bar{\lambda} \bar{\lambda}}{16\pi^2}\,G_3\wedge *\Omega +c.c.
\end{equation}
The action \eqref{llD7action} acts as a source for $G_3$ and a particular solution to the equations of motion is given by \cite{Dymarsky:2010mf}
\begin{equation}
\label{G3profile}
G^{\lambda\lambda}_3=G_3^++G_3^-\, ,
\end{equation}
with imaginary self-dual (ISD) component
\begin{equation}
e^{4\mathcal{A}}G_3^+=\frac{1}{\pi}e^{-\phi/2}\frac{\langle \lambda\lambda \rangle}{16\pi^2}(g^{i\bar{j}}\nabla_i \nabla_{\bar{j}} \Psi) \,\overline{\Omega}\, ,
\end{equation}
and imaginary anti-self-dual (IASD) component
\begin{equation}
(e^{4\mathcal{A}}G_3^-)_{i\bar{j}\bar{k}}=-\frac{i}{\pi} e^{-\phi/2}\frac{\langle \lambda\lambda \rangle}{16\pi^2}(\nabla_i \nabla_l \Psi)g^{l\bar{m}}\bar{\Omega}_{\bar{m}\bar{j}\bar{k}}\, ,
\end{equation}
where $\Psi$ is determined by $\nabla^2\Psi=2\pi\delta(\Sigma)$, and $\delta(\Sigma)$ is the scalar delta function that localizes on the four-cycle that the $7$-branes wrap\footnote{Strictly speaking, $\Psi$ is the solution to $\nabla^2 \Psi= 2\pi\left(\delta(\Sigma)-\frac{\text{Vol}(\Sigma)}{\mathcal{V}}\right)$ \cite{Baumann:2006th}. We will be interested in the behavior of supergravity fields in the vicinity of localized objects. In this regime the constant correction to the source of $\Psi$ only gives rise to small (i.e. volume suppressed) corrections which we shall neglect consistently.}. Note that $\Psi$ is identified with $\text{Re}\log h(z^i)$, where $h(z^i)=0$ is the holomorphic embedding equation of the $7$-brane divisor \cite{Baumann:2010sx}. The flux profile \eqref{G3profile} is only a particular solution to the equations of motion and is completed by the global harmonic fluxes.

Building on these results we now proceed. In order to match the $4D$ description of KKLT we should be able to determine the value of the volume modulus at the supersymmetric minimum from the value of the GVW superpotential. In order to do this in general one would have to derive the full off-shell potential \eqref{4dpotential_offshell} while we derive only the on-shell potential \eqref{4dpotentialIIB}. Fortunately if we assume a relation between the condensate and the $4$-cycle volume $\langle \lambda\lambda \rangle \sim e^{ia \rho}$ we may deduce the value of $\rho$ by demanding that the quantum deformed $10D$ SUSY conditions are fulfilled. These were derived in \cite{Dymarsky:2010mf} and we merely quote their result: to leading order in the gaugino condensate, the three-form flux $G_3$ is given by
\begin{equation}
\label{G3fluxprofile}
G_3=G_3^{(2,1)}+G_3^{\lambda\lambda}\, ,
\end{equation}
where $G_3^{(2,1)}$ is harmonic and of Hodge type $(2,1)$. Thus, the $(0,3)$ component of $G_3$ localizes completely on the $7$-brane divisor and is related to the value of the gaugino condensate via the $G_3$ equations of motion. As a consequence, just as in the $4D$ EFT description the value of the Gukov-Vafa-Witten superpotential determines the value of the condensate,
\begin{equation}
W_0\sim \int_{M_6} G_3\wedge \Omega\sim  e^{-\phi/2}\langle \lambda\lambda \rangle \sim e^{ia\rho}\, .
\end{equation}
Apart from the harmonic $(2,1)$ component the $G_3$ flux profile is thus fully determined. Furthermore one may show that the field $\Phi^-$ is not perturbed to leading order in the condensate. \par 
$\Delta^{loc}$ is given by
\begin{equation}
\label{deltalocexplicit}
\Delta_{loc}=-\frac{3\pi}{8} e^{\phi/2}e^{-4\mathcal{A}}\frac{\langle \bar{\lambda}\bar{\lambda} \rangle}{16\pi^2} \,\Omega\cdot G_3\, \delta^{(0)}_D+c.c.
\end{equation}
This means that all the ingredients of the on-shell potential are gathered and after some algebra (see appendix \ref{App:GauginoCondensateBackr.and.contr.to.onshell.cc}) we arrive at
\begin{equation}
\label{PotentialSingleCondensate}
V\cdot M_{\rm P}^{-4}=-\int_{M_6} \frac{d^6 y \sqrt{g}}{32\pi^2\mathcal{V}_w\tilde{\mathcal{V}}_w}(\alpha-\beta) \left|\frac{\langle \lambda\lambda \rangle}{16\pi^2}\nabla_i\nabla_j \Psi\right|^2\, .
\end{equation}
Here, the positive but otherwise unspecified numbers $\alpha,\beta$ are related to the fact that the integral in eq. \eqref{PotentialSingleCondensate} receives contributions of opposite sign from bulk-fluxes near the position of the seven-branes, as well as fluxes that are fully localized on the branes. Both contributions are UV-divergent
(see appendix \ref{App:GauginoCondensateBackr.and.contr.to.onshell.cc} for details)\footnote{We note that these divergences are very similar to the divergent self-energy of an anti-brane at the bottom of a warped throat which is explained in \cite{Michel:2014lva,Polchinski:2015bea}.}. We expect that imposing a string-scale cutoff gives the right result up to regulator dependent $\mathcal{O}(1)$ factors $\alpha,\beta$\footnote{In an earlier version of this paper, the stronger claim was made that the ratio $\alpha/\beta$ could be determined uniquely which was based on an ad-hoc regularization of UV-divergences. However, as pointed out to us by the authors of \cite{Gautason:2018toappear} this would be inconsistent with the assumption that supersymmetric KKLT vacua exist. Hence, the constraint $\alpha>\beta$ must be imposed by hand, rather than being a prediction of the $10D$ calculation itself.}. Assuming the existence of supersymmetric KKLT vacua we conclude that $\alpha>\beta$, so that the cosmological constant is negative and proportional to the strength of the gaugino condensate. \par 
One can readily generalize this expression to the case of $n$ stacks of $7$ branes that wrap different holomorphic representatives of the same divisor,
\begin{equation}
\label{PotentialMultipleCondensates}
V\cdot M_{\rm P}^{-4}=-\int \frac{d^6 y \sqrt{g}\, }{32\pi^2\mathcal{V}_w\tilde{\mathcal{V}}_w}\left(\alpha\left|\sum_{a=1}^{n}\frac{\langle \lambda\lambda \rangle_a}{16\pi^2}\nabla_i\nabla_j \Psi_a \right|^2-\beta\sum_{a=1}^{n}\left|\frac{\langle \lambda\lambda \rangle_a}{16\pi^2}\nabla_i\nabla_j \Psi_a \right|^2\right)\, .
\end{equation}
Crucially, the sign of the cosmological constant depends on the relative phases between the condensates $\langle \lambda\lambda \rangle_a$. This suggests that for two condensates one may be able to obtain a cosmological constant that is parametrically smaller than the strength of the individual condensates that set the mass-scale of the lightest modulus. For more than two condensates it may even be possible to obtain de Sitter solutions.\par 
This fits nicely with the effective $4D$ description of multiple condensates: the case of two condensates was studied by Kallosh and Linde \cite{Kallosh:2004yh} who show that it is possible to tune the cosmological constant to zero supersymmetrically while retaining a finite volume stabilization.\par
 
Before we proceed one more comment is in order: the integrals \eqref{PotentialSingleCondensate} and \eqref{PotentialMultipleCondensates} are quadratically UV-divergent. Imposing a UV-cutoff $\Lambda_{UV}$ one finds
\begin{equation}
|V|\cdot M_{\rm P}^{-4}\sim \frac{\text{Vol}(\Sigma_4)}{\mathcal{V}^2}\left|\frac{\langle \lambda\lambda \rangle}{16\pi^2} \right|^2\Lambda_{UV}^2+\text{finite}\, .
\end{equation}
Since we are not interested in overall order one coefficients we leave a proper EFT treatment of this divergence for future work and simply cut-off the integrals at the string scale\footnote{Recall that we work in $10D$ Einstein frame in units $M_{{\rm P},10D}^8=4\pi$.} $\Lambda_{UV}^2\sim e^{\phi/2}$.

\subsection{The uplift}\label{10duplift}

\begin{figure}[t!]
	\centering
	\includegraphics[keepaspectratio, width=8cm]{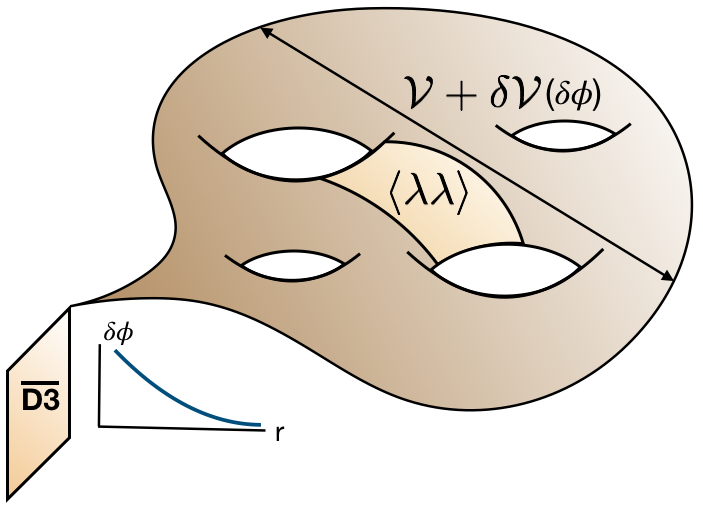}
	\caption{Two aspects of $\overline{D3}$ back-reaction: on the one hand the local throat perturbations sourced by the $\overline{D3}$-brane at the bottom of the throat fall off exponentially towards the bulk Calabi-Yau. Their UV-tail can be neglected in our analysis. On the other hand the overall volume of the bulk Calabi-Yau adjusts and so does the condensate $\langle \lambda \lambda \rangle$. This \textit{global} back-reaction effect turns out to be crucial for our analysis.}
	\label{fig:RadialThroatProfiles}
\end{figure}

In order to determine if an anti-brane (or in fact any other source of SUSY breaking) at the bottom of a warped throat can uplift to de Sitter in the presence of only a single condensate, we have to take a closer look at the on-shell potential \eqref{4dpotentialIIB}. The anti-brane perturbs it in two ways (see figure \ref{fig:RadialThroatProfiles}),
\begin{enumerate}
	\item It sources all local $10D$ supergravity fields in the throat. Away from the brane their field profiles are harmonic and fall off towards the bulk-CY.
	\item It pulls on the volume modulus, the lightest degree of freedom in the problem. Because we assume an exponential relation between the condensate and the volume modulus, the value of the condensate is changed as well.
\end{enumerate}
Let us focus on the local back-reaction first. The naive approach to include the effect of the $\overline{D3}$-brane would be to compute its corresponding contribution to the on-shell potential. The outcome is   $\Delta^{\text{loc}}_{\overline{D3}}=2T_3>0 $ ($T_3$ is the tension of the brane), so the inclusion of the $\overline{D3}$-brane implies a new \textit{negative} contribution to \eqref{4dpotentialIIB}. Then, how is the anti-brane  going to provide an uplift? As we already saw in the toy model, the use of on-shell potentials comes with certain peculiarities: The dominant source of uplift (the three-brane tension) did not appear \textit{explicitly} in the on-shell potential. Rather its uplifting effect is contained \textit{implicitly} in the induced shift in the volume modulus by which the uplift can be efficiently parametrized even when the back-reacted solution is not known. Now, we are in a similar situation. The explicit appearance of the anti-brane via $\Delta^{loc}$ and in fact all the details of the back-reacted warped throat geometry are suppressed by eight powers of the warp factor.\footnote{Note that close to the anti-D3 brane the potential $\Phi^-$ is sourced at ${\cal O}(e^{4\mathcal{A}_0})$. Hence, the integrand of the on-shell potential is dressed with an overall factor of $e^{8\mathcal{A}}$.} We know from the off-shell potential \eqref{4dpotential_offshell} that the leading order uplift should be of order $e^{4\mathcal{A}_0}\ T_3$, and therefore, if we only work to order $e^{4\mathcal{A}_0}$ we can completely neglect the local physics of the throat, making the sign of $\Delta^{\text{loc}}_{\overline{D3}}$ irrelevant. Instead, as in the toy example, the leading order uplift is parametrized implicitly by the induced back-reaction on the \textit{bulk} Calabi-Yau on which we will focus in the following.

Before doing so let us set a precision goal for the upcoming analysis. Because the minimal uplifting potential must compete with the $AdS$ depth of the SUSY KKLT minimum one should have $e^{4\mathcal{A}_0}\gtrsim |\langle \lambda \lambda \rangle |^2$ as a minimal requirement. Guided by this in the following we will work to leading order in
\begin{equation}
\label{expansion_scheme}
e^{(4-2q)\mathcal{A}_0}|\langle \lambda \lambda \rangle|^q\, ,
\end{equation}
for any $q$ and consistently neglect higher combined powers of the condensate and the warp factor.\par

The next step is to estimate how the local physics of the bulk is affected by the SUSY breaking source in the strongly warped region. Of course one cannot use on-shell methods to do this. But since the supergravity solution corresponding to a warped throat is known in full detail, this estimate can be done explicitly by perturbing the IR-end of the throat and deriving how fast the corresponding field profiles fall off towards the UV. Building on the work of \cite{Baumann:2010sx,Gandhi:2011id,GMS} we have provided a treatment of this question in appendix \ref{App:RadialProfiles}. 

The outcome is simple and intuitive: all field profiles fall off exponentially towards the bulk such that for an IR perturbation of order one the corresponding perturbation in the UV is suppressed by powers of the IR warp factor $e^{\mathcal{A}_0}$. If we neglect dependencies on the CY volume which we assume to be only moderately large it turns out that to leading order in $e^{\mathcal{A}_0}$ all \textit{local} supergravity fields receive corrections of order 
\begin{equation}
\mathcal{O}(e^{p\mathcal{A}_0}\cdot T_3)\, ,\quad p\in \{3,4\}\, .
\end{equation} 
Because both $G_3^-$ as well as $\Phi^-$ appear quadratically in \eqref{4dpotentialIIB} we may neglect their adjustment that is sourced directly by the anti-brane.  Also the localized contribution in \eqref{deltalocexplicit} contains an overall factor of $\langle \lambda \lambda \rangle$. Hence, local field profiles would have to be corrected at least at order $e^{2\mathcal{A}_0}$ to enter our discussion. No such field profiles can be sourced directly by the antibrane.\par 

Hence, we need only consider the adjustment of the universal K\"ahler modulus.  However, calculating the value of the volume modulus is not possible using our on-shell methods and we lack the required off-shell methods to explicitly determine back-reaction effects in the bulk. The only reason that we were able to deduce its value at the supersymmetric point and relate it to the value of the GVW superpotential was that the $10D$ SUSY conditions gave us enough constraints. At no point did we explicitly minimize an effective off-shell potential of the volume modulus.
	
Let us argue why we are able to proceed despite this shortcoming of the on-shell methods. First, the validity of the on-shell methods as such does not require supersymmetry. We assume that once the antibrane is included all fields adjust to their new respective minima. Hence once the system has found its energetically most favorable configuration the outcome must be consistent with the on-shell potential \eqref{4dpotentialIIB}. The only technical difference between the supersymmetric and non-supersymmetric configuration is that we are no longer able to determine the shifted value of the volume modulus. For our purposes this is not a problem. We simply assume that the new minimum lies at a different volume
\begin{equation}
\rho\longrightarrow \rho +\delta \rho\, ,
\end{equation}
 and express all the quantities that need to be evaluated as functions of the new (unknown) value of the volume modulus. When the volume modulus assumes a new value, all seven-brane sourced effects adjust accordingly. Discarding effects that are sub-leading in inverse volume (to be addressed shortly), this means that in configurations with one or more gaugino condensates stabilizing the K\"{a}hler modulus, upon the inclusion of the anti-brane, equations \eqref{PotentialSingleCondensate} and \eqref{PotentialMultipleCondensates} still hold albeit as a function of the (unknown) shifted value of the volume modulus.    Hence, from (\ref{PotentialSingleCondensate}) it is easy to see that in the case of a single gaugino condensate even in the presence of the anti-brane the on-shell potential is manifestly negative.

Ultimately, the reason that the contribution of a single gaugino condensate to $V$ is always negative rests on the fact that its contribution to the integrand of the on-shell potential is always negative\footnote{The reason for this is that the 7-brane contribution to $\Delta \equiv 2\pi\frac{|G_3^-|^2}{\text{Im}(\tau)}+\Delta^{loc}$ is positive.} (see appendix \ref{App:GauginoCondensateBackr.and.contr.to.onshell.cc} for details). Although there are two 7-brane induced contributions to the integrand of either sign from the induced bulk and brane-localized fluxes, the negative contribution of the 7-brane-induced bulk piece $G_3^-$ overcompensates the 7-brane-localized $G_3^+$ piece. This happens for all values of $\rho$ as long as both $G_3^+$ and $G_3^-$ are sourced by a unique 7-brane world volume coupling of the condensate $\langle\lambda\lambda\rangle$ to $G_3$. We have shown in appendix~\ref{App:GauginoCondensateBackr.and.contr.to.onshell.cc}, that in absence of any other fermion condensates besides $\langle\lambda\lambda\rangle$ the coupling $\langle\lambda\lambda\rangle G_3\cdot\Omega$ is unique at $\mathcal{O}(\langle\lambda\lambda\rangle)$\footnote{We use the straightforward non-abelian generalization of the single $D7$-brane outcome, guided by our expectation that gauge-symmetry and supersymmetry will uniquely fix this.}. This statement holds at the level of the 10D equations of motion and does not use background supersymmetry for the derivation.

Hence, in the end the situation here is analogous to the Freund-Rubin toy model (without the 5-branes) and also here it follows immediately that vacua that are stabilized by  a single condensing non-abelian gauge group (in a regime where perturbative corrections can be neglected, see section \ref{sec:alphaprime}) cannot be uplifted to de Sitter.  

As we have explained we cannot in principle derive the adjusted value of the volume modulus and the corresponding change in the cosmological constant from the on-shell potential. The on-shell potential \eqref{PotentialSingleCondensate} only reveals the overall sign of the resulting cosmological constant but does not allow to compute its value. For this one would need the full off-shell potential which we currently lack. Nevertheless, we can attempt a parametric estimate (neglecting volume powers) based on generic assumptions about the coupling of an uplift parametrized by the warped tension $e^{4\mathcal{A}_0}T_3$ to the volume modulus $\rho$ (see appendix \ref{App:RadialProfiles}) 
\begin{equation}
\delta \rho \sim \dfrac{e^{4\mathcal{A}_0}\ T_3}{    m_{\rho}^2} \sim \dfrac{e^{4\mathcal{A}_0}\cdot T_3 }{ |\langle \lambda\lambda \rangle|^2} \, .
\end{equation} 
This estimate is easy to understand: the weaker the stabilization the stronger the volume modulus will react to a small IR perturbation. Therefore for $e^{4\mathcal{A}_0} T_3 \ll |\langle \lambda \lambda \rangle|^2$ we can plug this estimate into the on-shell potential \eqref{PotentialSingleCondensate} to obtain
\begin{equation}
\delta V\sim \underbrace{V}_{\sim |\langle \lambda\lambda \rangle|^2}\cdot \underbrace{\delta \rho}_{\sim e^{4\mathcal{A}_0}\cdot T_3/|\langle \lambda\lambda \rangle|^2} \sim e^{4\mathcal{A}_0}\cdot T_3\, ,
\end{equation} 
which describes the naive result of adding energy contributions. Note that this regime corresponds precisely to the region of parametric control where the uplift is small compared to the initial vacuum energy $\delta V \ll |V_{AdS}|$ and thus the final state is also AdS.  

When the warped tension is not much smaller then $|V_{AdS}|$ it is not possible even to estimate the magnitude of the uplift since  $\delta \rho$ is no longer small. From the on-shell potential we only know that the final state will be $4D$ $AdS$ and the inclusion of the anti-brane should increase the potential energy. From here we conclude that its magnitude will be flattened out due to strong backreaction on the volume modulus and thus the final energy is not just the sum of $V_{AdS}$ and the warped anti-brane tension.  
 
\begin{figure}[t!]
\begin{center}
	\includegraphics[keepaspectratio, width=10cm]{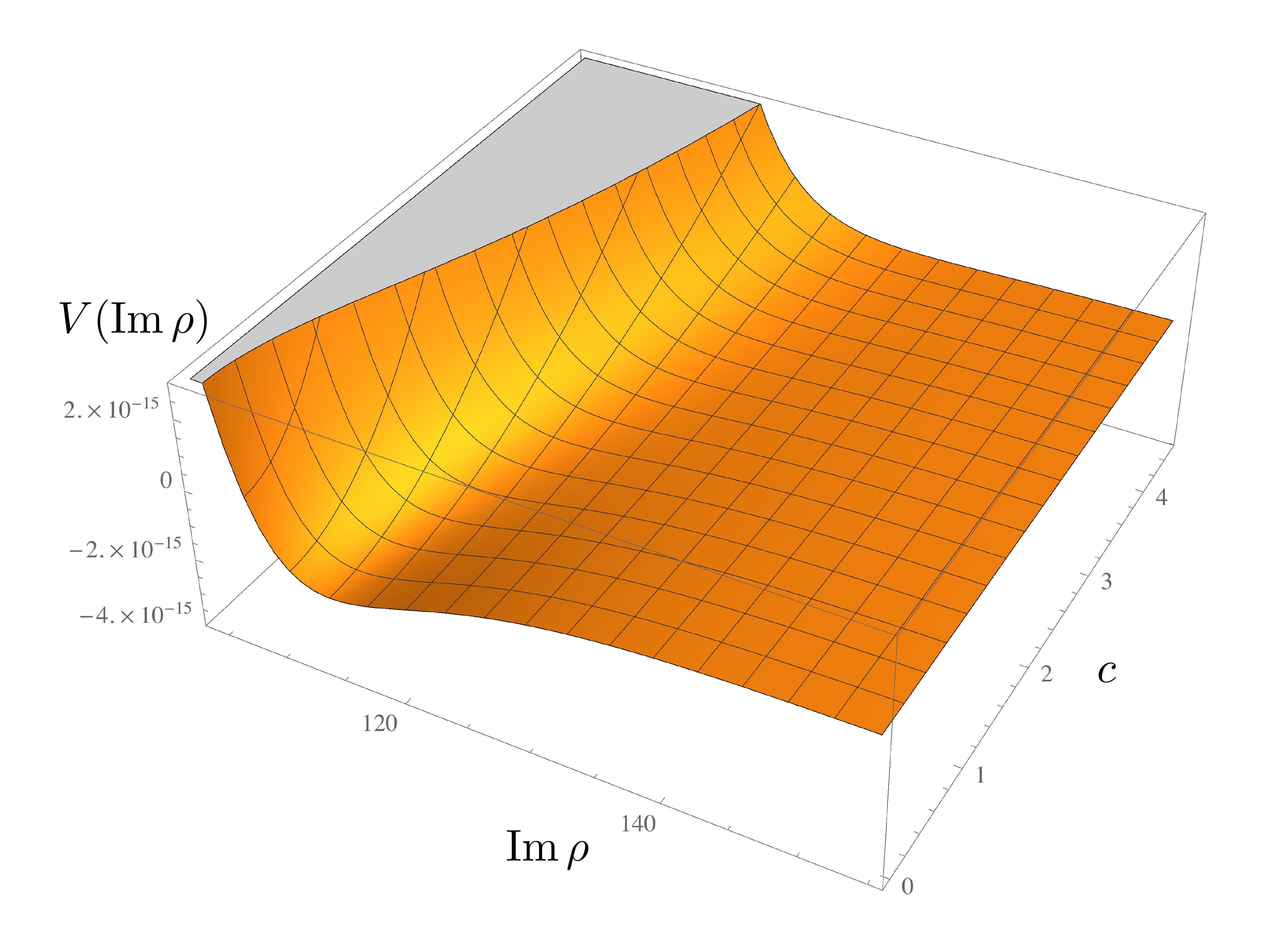}
\end{center}
	\caption{The F-term potential of \eqref{antibranesuperpot} for the case $b=0$ as a function of the uplift parameter $c$. In this case the superpotential reads $W=W_0+A\,(1+c\, S) \exp(ia\rho)$. The $\rho$-minimum moves out to larger values and stays $AdS$ as $c$ increases. This behavior is compatible with our $10D$ analysis but does not uniquely follow from it.}
	\label{fig:money.plot}
\end{figure}

It is now interesting to go back to  the $4D$ superpotential \eqref{antibranesuperpot} to note that our $10D$ analysis is compatible with the extreme case $b \rightarrow 0$, suggesting the existence and significant strength of the superpotential term describing the interaction between the anti-brane and the gaugino condensate (see figure \ref{fig:money.plot}). We will comment further on this in section \ref{10Dinterpret}.

%%%%%%%%%%%%%%%%%%%%%%%%%%%%%%%%%%%%%%%%%%%%%%%%%%%%%%%%%%%%%%%%%%%%%%%%%%%%%%%%%%%%%%
%%%%%%%%%%%%%%%%%%%%%%%%%%%%%%%%%%%%%%%%%%%%%%%%%%%%%%%%%%%%%%%%%%%%%%%%%%%%%%%%%%%%%%

For completeness, we would like to comment here on an interesting effect of the volume shift caused by the $\overline{D3}$-brane.  We are assuming an exponential relation between the condensate $\langle \lambda\lambda \rangle$ and the volume modulus, and so the condensate must be reduced in magnitude due to the shift. Since the profile $G_3^{\lambda\lambda}$ is uniquely determined by the strength of the gaugino condensate, the situation is so far analogous to the supersymmetric case except that we have to allow for a global harmonic $(0,3)$ component in the three-form
\begin{equation}
G_3=G_3^{(2,1)}+G_3^{(0,3)}+G_3^{(\lambda\lambda)}\, .
\end{equation}
The global $(0,3)$ component enters the on-shell potential \eqref{4dpotentialIIB} via $\Delta^{loc}$ and thus gives rise to a further term in the on-shell potential. The magnitude of the global $(0,3)$ component is determined \textit{as a function of the shifted volume} by demanding the value of $\int G_3 \wedge \Omega$ to be conserved.\footnote{Here we assume that back-reaction on complex structure moduli is negligible.} Since the localized $(0,3)$ piece is reduced in comparison with the supersymmetric setup, part of it is converted into the global $(0,3)$ component. This harmonic $(0,3)$ component is interesting because it determines soft masses on branes \cite{Camara:2003ku,Grana:2003ek,Camara:2004jj} of order the warped tension of the anti-brane. For our purposes, it enters the on-shell potential \eqref{4dpotentialIIB} via $\Delta^{loc}$ only with a relative volume suppression with respect to e.g. the localized flux contribution on the D7-brane stack (see appendix \ref{App:vol.supp.contr.softmasses} for details). We work to leading order in $1/\mathcal{V}$ and  therefore neglect this contribution.

%%%%%%%%%%%%%%%%%%%%%%%%%%%%%%%%%%%%%%%%%%%%%%%%%%%%%%%%%%%%%%%%%%%%%%%%%%%%%%%%%%%%%%

\begin{figure}[t!]
\begin{center}
	\includegraphics[keepaspectratio, width=10cm]{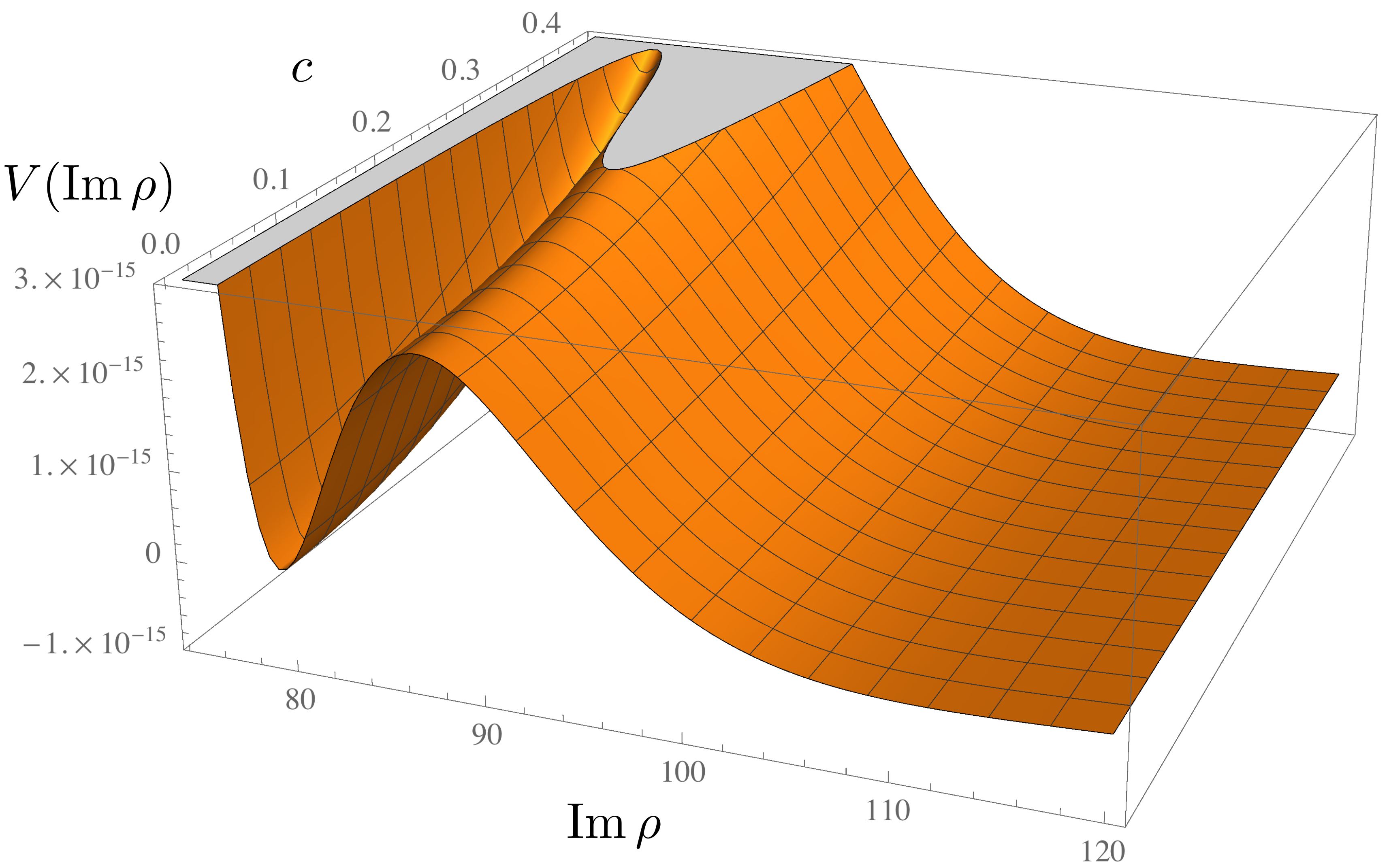}
\end{center}
	\caption{The F-term potential analogous to \eqref{antibranesuperpot} for the case $b=0$ as a function of the uplift parameter $c$ for racetrack superpotential. In this case the superpotential reads $W=W_0+A_1\,(1+c\, S) \exp(ia_1\rho)+A_2\,(1+c\, S) \exp(ia_2\rho)$. For simplicity we have put the coefficients of the $S$-dependence in both gaugino condensates equal. For vanishing uplift $c=0$ the scalar potential shows the tuned racetrack Minkowski minimum at smaller values of $\rho$ then the co-existing KKLT $AdS$ minimum at larger values or $\rho$. Clearly, the racetrack minimum gets successfully uplifted to de Sitter even for $c\ll 1$ while the KKLT-minimum at larger $\rho$ continues to move out to larger values and stays $AdS$ as $c$ increases. Note, that this a 4D effective extrapolation to the two-condensate case, for which we do not have a fully matching 10D description yet.}
	\label{fig:money.plot.racetrack}
\end{figure}

The obstruction towards reaching de Sitter in the single condensate configuration can  be evaded in a rather simple generalization of the setup: in racetrack configurations  with at least two condensing non-abelian gauge groups the multi condensate potential \eqref{PotentialMultipleCondensates} confirms from a $10D$ perspective the possibility to decouple the lightest modulus mass from the $AdS$-depth in the supersymmetric vacuum by giving the individual condensates opposite phases. In a limiting case it may even allow for Minkowski vacua after stabilizing the K\"{a}hler modulus. This situation would correspond to the case when the two terms in \eqref{PotentialMultipleCondensates} are of the same magnitude upon integration and thus compensate each other. Note that in this case there exists the required positive contribution to the on-shell potential, allowing to evade the Maldacena-N\'{u}\~{n}ez theorem, in analogy with the Freund-Rubin toy model when the 5-brane is  present. Therefore, the inclusion of the $\overline{D3}$-brane on the bottom of a warped  throat in this configuration could well provide the necessary uplift to de Sitter (see Figure \ref{fig:money.plot.racetrack}).  

While we cannot prove the full $10D$ consistency of such racetrack stabilization mechanisms we see no sign of failure in an anticipated fully fledged back-reacted $10D$ solution. If the required tuning can be realized it would seem that such vacua can be uplifted to de Sitter \textit{generically} (as explained in section \ref{KKLT:uplift}), i.e. by \textit{any} SUSY breaking energy density that is red-shifted sufficiently strongly to suppress back-reaction on the volume modulus (i.e. $e^{4\mathcal{A}_0}\ll |\langle\lambda\lambda\rangle_a|^2$). Thus, it seems that such two-condensate (racetrack) stabilization mechanisms are ideal backgrounds for uplifting scenarios, independently of the details of the uplifting mechanism, and the details of its couplings to the lightest moduli. However, we would also like to point out that the mere existence of warped throats with warping in the regime $e^{4\mathcal{A}_0}\ll |\langle\lambda\lambda\rangle_a|^2$ is not obvious.

\subsection{Towards an interpretation of an unsuppressed $S\langle \lambda\lambda\rangle$ coupling}
\label{10Dinterpret}
We would like to conclude this section with an attempt of a physical interpretation of our result. We have constrained the anticipated fully fledged backreacted solution using on-shell methods. Although these methods are powerful enough to constrain the sign of the cosmological constant they reveal very little about its physical origin. So, let us speculate about it:\par
We are assuming that near the bottom of the warped throat the local description of a warped throat à la Klebanov-Strassler still holds, while at the same time we believe that the $4D$ EFT description of the anti-brane state using a nilpotent superfield is valid. Putting these two expectations together, and recalling the superpotential of eq. \eqref{antibranesuperpot1}, we are led to the interpretation that the IR warp factor itself receives corrections of order $|\langle \lambda\lambda \rangle|$, i.e.
\begin{equation}
e^{2\mathcal{A}_0}\longrightarrow e^{2\mathcal{A}_0}+\mathcal{O}(|\langle \lambda\lambda \rangle|)\, .
\end{equation}
We have asked ourselves if this expectation is reasonable in view of considerable effort that has been put into determining the position moduli potential of $D3$-branes \cite{Baumann:2010sx} and $\overline{D3}$-branes \cite{GMS}. In principle, the quantum corrected warp factor can be reconstructed by adding the $D3$- and $\overline{D3}$ potentials. In \cite{Baumann:2010sx} it was found that the $D3$-brane moduli potential as induced by the non-perturbative bulk effects is of order $e^{\mathcal{A}_{cl,0}}|\langle \lambda\lambda \rangle|^2$ and is hence suppressed by a classical warp factor. However, to the best of our knowledge, the tools used in \cite{Baumann:2010sx} are only sensitive to the \textit{force} that acts on the $D3$-brane, that is only the \textit{position-dependent} part of the $D3$-brane moduli potential is determined. Their results are hence compatible with the correction that we have suggested.\par
If this correction indeed occurred, an interesting consequence would be that the IR warp factor receives significant corrections in the regime $e^{4\mathcal{A}_0}\sim |\langle \lambda\lambda \rangle|^2$. Nevertheless, the \textit{local} description of the IR-region of the warped throat would stay intact as long as the weaker requirement $|\langle \lambda\lambda \rangle|^2\ll e^{3\mathcal{A}_0}$ is fulfilled.\footnote{If this requirement is violated local IASD field strengths would become dominant \cite{Baumann:2010sx}.} A local throat observer would notice the effect we are suggesting merely as a change of Newton's constant.\par 
An interesting prospect for future research would be to validate or falsify our interpretation by finding the explicit back-reacted warped throat solution. The tools needed to do this have been laid out in \cite{Gandhi:2011id} who have put the type IIB supergravity equations of motion into 'triangular' form suitable for determining the response of the throat to a source term in a systematic fashion. Following their strategy one could expand the fields $\Phi^-$ and $\Phi^+$ that determine the backreacted warp factor as
\begin{equation}
\Phi^-=\underbrace{\Phi^-|_{ISD}}_{=0}+\delta\Phi^-|_{\mathcal{IH}}+\delta\Phi^-|_{\mathcal{H}}\, ,
\end{equation} 
\begin{equation}
(\Phi^+)^{-1}=\underbrace{(\Phi^+)^{-1}|_{ISD}}_{=\frac{1}{2}e^{-4A}|_{ISD}}+\delta(\Phi^+)^{-1}|_{\mathcal{IH}}+\delta(\Phi^+)^{-1}|_{\mathcal{H}}\, ,
\end{equation}
where the subscripts $\mathcal{IH}$ and $\mathcal{H}$ denote the inhomogeneous respectively homogeneous part. Integrating the equations of motion would determine $\Phi^-$ as well as $\Phi^+$ from which the warp factor can be determined \textit{uniquely}.
%%%%%%%%%%%%%%%%%%%%%%%%%%%%%%%%%%%%%%%%%%%%%%%%%%%%%%%%%%%%%%%%%%%%
\section{Applying the lessons:\\ dS in KKLT with $\alpha'$-corrections?}\label{sec:alphaprime}

Let us use the lessons learned from the previous analysis to speculate about alternatives to the racetrack setup in the language of $4D$ EFT. From the type IIB analysis in~\cite{Becker:2002nn,Bonetti:2016dqh} as well as more recent lifts to F-theory in~\cite{Minasian:2015bxa} we know that the $R^4$-terms and their supersymmetric completion in 10D induce the leading correction to the 2-derivative kinetic terms of the K\"ahler moduli at ${\cal O}(\alpha'^3)$. 

The results of~\cite{Conlon:2005ki} demonstrated that other $\alpha'$-corrections involving higher powers of the RR and/or NSNS $p$-form field strengths produced 4D contributions suppressed by additional inverse powers of the compactification volume, while the results of~\cite{vonGersdorff:2005bf,Berg:2005ja,Berg:2007wt,Cicoli:2007xp,Cicoli:2008va,Berg:2014ama} show that the same is true for string loop corrections due to the extended no-scale structure present in type IIB Calabi-Yau compactifications. As a result, the K\"ahler potential of the volume moduli acquires at leading order an ${\cal O}(\alpha'^3)$ correction
\begin{equation}
K=-2\ln(\mathcal{V}+\xi/2)+K_{c.s.}\quad{\rm with:}\quad \xi \sim - g_s^{-3/2} \chi_{CY}
\end{equation}
where $\chi_{CY}=2 (h^{11}-h^{21})$ is the Euler characteristic of the Calabi-Yau in question.

\begin{figure}[t!]
	\centering
	\includegraphics[keepaspectratio, width=97mm]{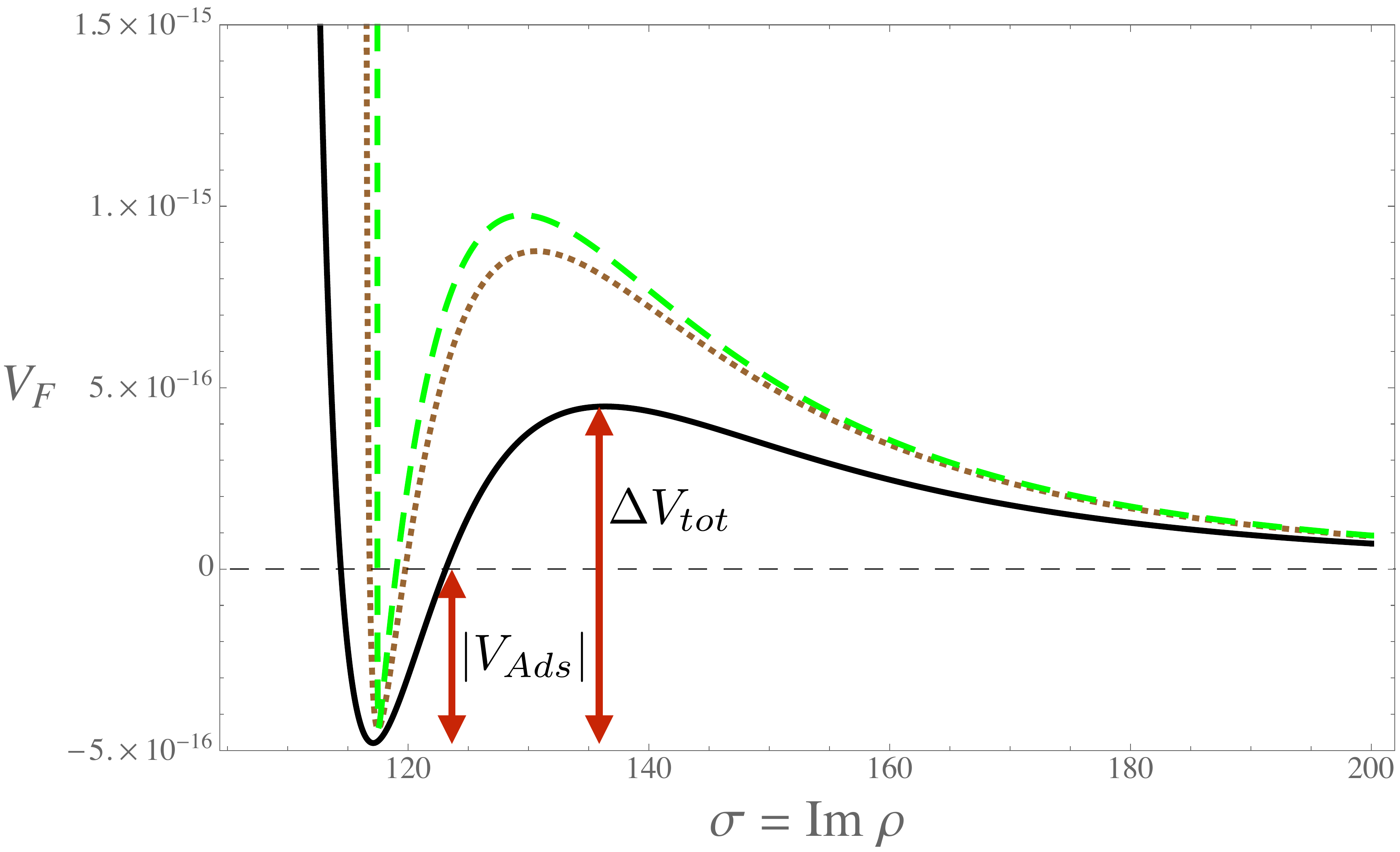}
	\caption{Volume stabilization in KKLT with a single gaugino condensate but including the leading ${\cal O}(\alpha'^3)$-correction to the volume moduli effective action. For $\xi\to \mathcal{V}$ we see that the correction begins to create a barrier towards decompactification at positive values of the potential. We see that the total potential difference between the SUSY minimum and the barrier top can become as large as about $4\times |V_{AdS}|$. The curves show the potential for values of $\xi=3.1\times 10^3$ (black solid), $\xi=3.55\times 10^3$ (brown dotted), and $\xi=3.6\times 10^3$ (green dashed). }
	\label{fig:alphaKKLT}
\end{figure}

We again assume supersymmetric flux stabilization of the c.s. moduli and axio-dilaton. We then proceed to use the superpotential of KKLT $W=W_0+A e^{ia\rho}$. The resulting scalar potential is (see \cite{Becker:2002nn,Balasubramanian:2005zx})

\begin{eqnarray}
V_F&=&e^K \left[K^{\rho\bar\rho}\left(a^2 A^2 e^{ia (\rho-\bar\rho)}+iaA\left(e^{ia\rho} \bar W K_{\bar\rho}-e^{-ia\bar\rho} W K_{\rho}\right)\right)+ 3\xi \frac{\xi^2+7\xi\mathcal{V}+\mathcal{V}^2}{(\mathcal{V}-\xi) (2\mathcal{V}+\xi)^2}|W|^2\right]\nonumber \\ &&\nonumber\\
&=&\frac{e^{K_{c.s.}}}{8\, \sigma^3} \left[\frac{4  \sigma^2}{3}\left(a^2A^2 e^{-2a\, \sigma}+2aA e^{-a\, \sigma}\,{\rm Im}\left(e^{-ia{\rm Re}\,\rho}W K_\rho\right)\right) \right.\nonumber\\
&&\qquad\qquad\qquad\qquad\qquad\qquad \qquad\qquad \left. + 3\xi \frac{\xi^2+7\xi \sigma^{3/2}+ \sigma^3}{( \sigma^{3/2}-\xi) (2 \sigma^{3/2}+\xi)^2}|W|^2\right]
\end{eqnarray}
where we defined $\sigma \equiv {\rm Im}\,\rho = \frac12\mathcal{V}^{2/3}$. We plot this potential as a function of $\sigma$ for the parameter choice of the original numerical example in KKLT: $a=0.1$, $A=1$, $W_0=-10^{-4}$ in figure~\ref{fig:alphaKKLT}. We see that the $\alpha'$-correction only starts to modify the vacuum energy of the SUSY KKLT vacuum significantly once $\xi$ is large enough to violate $\xi\ll\mathcal{V}$.

Looking at the figure, we observe the following. Firstly, $m_\rho$ is determined roughly by the whole potential difference between the minimum and the barrier top. Secondly, by arranging for $\xi\to\mathcal{V}$ we can get a finite hierarchy between the AdS vacuum energy of the minimum and the total potential difference between the minimum and the barrier top $|V_{AdS}|\gtrsim 1/3 \Delta V_{tot}< \Delta V_{tot}$ where $\Delta V_{tot}=V_{barrier}-V_{top}$. Putting this together, this implies that we can arrange for a finite hierarchy of the necessary dS uplift  $\delta V\sim |V_{AdS}|\sim 1/3 \Delta V_{tot}<\Delta V_{tot}\sim m_\rho^2M_{\rm P}^2$.

 As this hierarchy was the condition for perturbatively small uplifting to de Sitter which we learned from the preceding discussion, it is possible that this simple modification of KKLT with a single gaugino condensate for each volume modulus might allow for  perturbatively controlled uplifts to de Sitter. Note however, that for the hierarchy $m_\rho^2M_{\rm P}^2 > |V_{AdS}|$ to emerge, we need to have $\xi\simeq \mathcal{V}$. This is a regime where we might worry about further, higher $\alpha'$-corrections no longer being smaller than the leading one, so further analysis is necessary to clarify the viability of this modification of KKLT to get to de Sitter. 

Given these observations, it is tempting to speculate about the situation in models where the $\alpha'$-correction to $K$ is under parametrical control, such as the Large Volume Scenario (LVS)~\cite{Balasubramanian:2005zx}. In its simplest version it involves a CY with two K\"ahler moduli $\rho_b$ and $\rho_s$ and a volume of swiss cheese form
	\begin{equation}
	\mathcal{V}= \sigma_b^{3/2}-\sigma_s^{3/2}
	\end{equation}
	where we denote the 4-cycle volumes as $\sigma_i = {\rm Im}\,\rho_i$. Furthermore, the only relevant non-perturbative contribution to $W$ will wrap $\sigma_s$
	\begin{equation}
	W= W_0+A_s e^{i a_s\rho_s}\quad.
	\end{equation}
	For this setup there exists for generic values of $|W_0|\sim \mathcal{O}(1)$ a minimum for both $\mathcal{V}$ and $\sigma_s\sim \xi^{2/3}\sim \mathcal{O}(10)$ at exponentially large volume
	\begin{equation}
	\mathcal{V} \sim |W_0| e^{a_s \sigma_s}\gg \sigma_s^{3/2} \quad.
	\end{equation}
	The interesting observation for our purpose here is, that this minimum is a SUSY breaking AdS minimum with the following scales
	\begin{equation}
	|V_{AdS}|\sim \frac{|W_0|^2}{\mathcal{V}^3} \sim m_{\cal V}^2 M_{\rm P}^2\sim m_{\sigma_b}^2 M_{\rm P}^2 \quad\ll\quad m_{\sigma_s}^2 M_{\rm P}^2 \sim \frac{|W_0|^2}{\mathcal{V}^2} \quad.
	\end{equation}
	Hence, in the LVS setup the amount of anti-brane uplift $\delta V \sim |V_{AdS}| \sim \mathcal{V}^{-3} \ll m_{\sigma_s}^2 M_{\rm P}^2$ is much smaller than the mass scale of the only 4-cycle volume directly controlling the scale of a non-perturbative effect.

Naively, this seems to satisfy the upshot of our preceding 4D and 10D analysis, namely the controlled-uplift condition $\delta V \ll m_{\phi_i}^2 M_{\rm P}^2$ for all moduli $\phi_i$ involved in non-perturbative moduli stabilization dynamics. However, LVS setups involve always at least two volume moduli, and their mechanism use the $\alpha'$-correction to produce one volume modulus with exponentially enhanced VEV via an exponential lever arm in terms of a small, but heavy blow-up volume modulus. At this point it is therefore far from clear, that the naive scale separation between the necessary amount of uplift in LVS models and the parametrically heavier blow-up 4-cycle controlling the leading non-perturbative effect, will shield the light overall volume direction sufficiently from the backreaction effects of uplift to conclude a higher level of safety of LVS compared to KKLT models with a single non-perturbative effect per 4-cycle.

Another important point pertains to the fact, that the inclusion of the $\alpha'$-correction is so far only possible in 4D EFT, given that its effect on the scalar potential only emerges after compactification. It is an open question how to include such $\alpha'$-corrections from higher-curvature terms at the level of the 10D description directly. Hence, we leave the analysis of the 10D stability of scenarios like LVS under uplifting as a tantalizing problem for the future.

%%%%%%%%%%%%%%%%%%%%%%%%%%%%%%%%%%%%%%%%%%%%%%%%%%%%%%%%%%%%%%%%%%%%%%%%%%

%%%%%%%%%%%%%%%%%%%%%%%%%%%%%%%%%%%%%%%%%%%%%%%%%%%%%%%%%%%%%%%%%%%%%%%%%%
\section{Conclusion}\label{Concl}
%%%%%%%%%%%%%%%%%%%%%%%%%%%%%%%%%%%%%%%%%%%%%%%%%%%%%%%%%%%%%%%%%%%%%%%%%%

In this paper we have reported progress towards a 10D description of the steps involving volume moduli stabilization without and with anti-D3-brane uplifting as part of the KKLT mechanism of generating a landscape of controlled meta-stable de Sitter vacua in type IIB string theory. We started by outlining a simple bottom-up 4D EFT argument indicating that models which require an uplift of the order $\delta V\gtrsim m_\rho^2M_{\rm P}^2$ (such as the simplest version of KKLT) are highly sensitive to the functional form of $\delta V(\rho)$ with $\rho$ being the lightest modulus. We have argued that with our current knowledge of non-perturbative volume stabilization and $\overline{D3}$-brane potentials the existence of SUSY breaking \textit{de Sitter} vacua within the simplest version of KKLT is hard to establish. 

It is for this reason that we have turned to a $10D$ description of the required non-perturbative effects that is powerful enough to decide the fate of the original version of the KKLT uplift as well as its racetrack improvement.

In order to do so, we have relied on two central methodological steps. Firstly,  we were interested in the \emph{energetics} of the volume modulus in a full 10D description. Assuming a solution to the full 10D system exists, we were interested in the \emph{on-shell} potential of this solution, and its scaling under changes of the stabilized CY volume. Conveniently this on-shell scalar potential incorporates all geometric back-reaction effects by expressing them via the matter contributions to the theory -- \emph{provided the backreacted solution exists}. For this purpose it was crucial that we could determine this on-shell potential from the matter sources alone by combining the \emph{trace-reversed} 10D Einstein equations with the 10D Bianchi identities~\cite{Baumann:2010sx,Dymarsky:2010mf,Heidenreich:2010ad}.

We have analyzed a simple toy model in 6D consisting of quantized 2-form flux on a positively curved $S^2$ and 3-branes, where fully backreacted solutions are easy to establish. We showed that sufficient uplifting by the 3-branes to naively reach de Sitter led to significant \emph{flattening} of the uplift: by this we mean that the increase of the $S^2$ radius $R$ with increasing uplift occurred such that the total vacuum energy remained AdS while approaching zero from below, supporting the 4D EFT worry. Without including the backreaction on $R$ the uplift would have been linear in the tension of the 3-branes which would have na\"ively allowed for de Sitter minima.

We then extended this toy model by adding a positive 6D cosmological constant. This allowed us to somewhat decouple the scale of the AdS vacuum energy from $m_\rho^2$ prior to adding the 3-branes for uplifting. In this case, the flattening from volume increase becomes less pronounced, and it is possible to reach de Sitter. From here we concluded that for models where reaching de Sitter requires $\delta V_{\overline{D3}}\gtrsim m_\rho^2M_{\rm P}^2$ according to the \emph{off-shell} potential, the details of the method of volume stabilization are essential.

Based on this, we needed to re-analyze volume stabilization in KKLT itself in 10D. We started by re-deriving the back-reaction onto the flux geometry caused by the gaugino condensation process on a D7-brane stack in the KKLT setup based on~\cite{Baumann:2010sx,Dymarsky:2010mf,Heidenreich:2010ad} and extracting for the first time the full contribution from the D7-gaugino condensate to the matter field side of the \emph{trace-reversed} Einstein equations. This allowed us to calculate an \emph{on-shell} potential for the $\rho$-modulus in the simplest KKLT setup. In turn, this enabled us to understand the minimization of the $\rho$-potential into a SUSY AdS minimum while matching the step 2 4D effective description outlined above. A crucial result of this analysis is that the $(0,3)$ component of the 3-form flux generating the small $|W_0|$ VEV of the Gukov-Vafa-Witten flux superpotential in step 1 becomes effectively localized at the D7-brane stack due to the backreaction to the formation of the gaugino condensation on the 7-branes.

Finally, we added back an anti-D3-brane at the bottom of the throat to complete the KKLT recipe. We observed that the in general complicated field profiles that the anti-brane sources in the warped throat fall off too rapidly towards the bulk Calabi-Yau to cause relevant effects in the on-shell potential. In contrast the on-shell potential preserves its functional form but is to be evaluated at the \textit{adjusted} value of the Calabi-Yau volume and the condensate. In other words we have calculated the \emph{on-shell} potential for $\rho$ as a function of the anti-D3 induced $\rho$-increase.

Because for a \emph{single} gaugino condensate arising from a single D7-brane stack the on-shell potential remains manifestly negative (at leading order in an expansion in inverse compactification volume) the 4D vacuum energy \emph{flattens} out as a function of increasing anti-D3-brane tension sufficiently strongly to prevent a successful uplift to de Sitter. The simplest KKLT vacua are therefore meta-stable SUSY breaking AdS vacua always. For sufficiently small warped anti-brane tension the uplift simply adds to the $AdS$ vacuum energy as predicted by the usual off-shell potential. We match this behavior to the effective 4D description of the anti-D3-brane by a nilpotent superfield $S$ appearing in the superpotential. The flattening of the uplift then requires the presence of an unsuppressed $S\langle\lambda\lambda\rangle$ coupling in the superpotential.

However, we then proceed to replace the single non-perturbative effect driving SUSY stabilization of $\rho$ by a \emph{racetrack superpotential} arising from a product gauge group with at least \emph{two} condensing subgroups (on one or several D7-brane stacks). In this case we can tune the supersymmetric AdS vacuum energy before uplifting arbitrarily close to zero, thus breaking the single-condensate KKLT relation $m_\rho^2M_{\rm P}^2\sim |V_{AdS}|$ into $m_\rho^2M_{\rm P}^2\gg |V_{AdS}|$. While this tuning possibility was long known in the context of the effective 4D supergravity description of the gaugino condensates, we provide a 10D description of this situation, and identify the 10D origin of the tuning possibility.

Relying on this tuning, the amount of uplifting $\delta V\sim |V_{AdS}|\ll m_\rho^2 M_{\rm P}^2$ is now perturbatively small. We can thus treat $\rho$ as rigid, which seems to make uplifting to de Sitter possible in the \emph{racetrack} case both from the 4D and 10D perspective.

Summing up our results, we believe that they enable a 10D understanding of the KKLT mechanism including its racetrack improvement. This relaxes the necessity of relying on the purely 4D methods employed so far for the steps of volume stabilization and uplifting in the KKLT framework.

\section*{Acknowledgements}

\par 
We thank  Wilfried Buchm\"uller, Markus Dierigl, Ben Freivogel, Fridrik Freyr Gautason, Vincent Van Hemelryck, Luis Ib\'{a}\~{n}ez, Fernando Marchesano, Paul Oehlmann,  Susha\\ Parameswaran, Thomas Van Riet, Irene Valenzuela, Timo Weigand, and especially Francisco Pedro, Eva Silverstein and Angel Uranga for various illuminating discussions. We are deeply grateful to Arthur Hebecker and Liam McAllister for carefully reading the manuscript and providing many very useful comments and extensive discussions. We are grateful to the SITP in Stanford, the IFT UAM/CSIC, and the PASCOS 2017 conference in Madrid, as well as the Lorentz Center Workshop ``Theoretical Approaches to Cosmic Acceleration'' in Leiden for their warm hospitality during various stages of this work. This work is supported by the ERC Consolidator Grant STRINGFLATION under the HORIZON 2020 grant agreement no. 647995.

%%%%%%%%%%%%%%%%%%%%%%%%%%%%%%%%%%%%%%%%%%%%%%%%%%%%%%%%%%%%%%%%%%%%%%%%%%
%%%%%%%%%%%%%%%%%%%%%%%%%%%%%%%%%%%%%%%%%%%%%%%%%%%%%%%%%%%%%%%%%%%%%%%%%%
\appendix
%%%%%%%%%%%%%%%%%%%%%%%%%%%%%%%%%%%%%%%%%%%%%%%%%%%%%%%%%%%%%%%%%%%%%%%%%%
\section{Type IIB and the No-go theorem of GKP}\label{App:TypeIIB}
%%%%%%%%%%%%%%%%%%%%%%%%%%%%%%%%%%%%%%%%%%%%%%%%%%%%%%%%%%%%%%%%%%%%%%%%%%

For flux compactifications of type IIB string theory the no-go theorem of \cite{Maldacena:2000mw} was extended significantly by Giddings, Kachru and Polchinski (GKP) \cite{Giddings:2001yu}: following them, we start with the Einstein frame action of type IIB Supergravity, 
\begin{equation}
\label{IIBaction}
S_{IIB}=2\pi \int_{M_{10}}\left(*R_{10}-\frac{d\tau\wedge *d\overline{\tau}}{2(\text{Im}(\tau))^2}-\frac{G_3\wedge *\overline{G_3}}{2\text{Im}(\tau)}-\frac{F_5\wedge *F_5}{4}-\frac{iC_4\wedge G_3\wedge\overline{G_3}}{4\text{Im}(\tau)}\right)+S_{loc}\, ,
\end{equation}
where $\tau=C_0+ie^{-\Phi}$ is the axio-dilaton, $G_3=F_3-\tau H_3$ is the complexified three-form field-strength, and 
\begin{equation}
F_5=dC_4-\frac{1}{2}C_2\wedge H_3+\frac{1}{2}B_2\wedge F_3\, .
\end{equation}
$S^{loc}$ is the action of localized objects and we have set $(2\pi)^2 \alpha'=1$.\par
Let us take the following ansatz for the $10D$ metric
\begin{equation}
ds^2=e^{2\mathcal{A}(y)}\tilde{g}^{(4)}_{\mu\nu}(x)dx^{\mu}dx^{\nu}+\underbrace{e^{-2\mathcal{A}(y)}\tilde{g}^{(6)}_{mn}(y)}_{\equiv g^6_{mn}(y)}dy^mdy^n\, ,
\end{equation}
where $\tilde{g}^4_{\mu\nu}$ is a maximally symmetric $4D$ metric. Then, the most general ansatz for $F_5$ that respects the symmetries of the $4D$ space-time is
\begin{equation}
F_5=dC_4+\mathcal{F}_5\, ,
\end{equation}
where $\mathcal{F}_5$ is purely internal and satisfies 
\begin{equation}
\label{F5Bianchi}
d\mathcal{F}_5=\frac{iG_3\wedge \overline{G_3}}{2\text{Im}(\tau)}+\frac{T_3}{2\pi}\rho_{D3}^{loc} \, ,
\end{equation}
with $D3$-brane charge $T_3=2\pi$ and $D3$-charge density of localized objects $\rho_{D3}^{loc}$. $C_4$ can be expressed as
\begin{equation}
\label{C4ansatz}
C_4=\alpha(y)\sqrt{-\tilde{g}^4}\,dx^0\wedge dx^1\wedge dx^2\wedge dx^3\, ,
\end{equation}
with a real function $\alpha(y)$ and is subject to the self-duality constraint $*dC_4=\mathcal{F}_5$.\par 
The $4D$ components of the $10D$ Ricci tensor can be expressed as
\begin{equation}
R_{\mu\nu}=\tilde{R}_{\mu\nu}-\frac{1}{4}g_{\mu\nu}e^{-4\mathcal{A}}\nabla^2e^{4\mathcal{A}}\, ,
\end{equation}
where $\tilde{R}_{\mu\nu}$ is the $4D$ Ricci tensor of the metric $\tilde{g}^4_{\mu\nu}$. Therefore,
the trace over the $4D$ components of Einsteins equations can be expressed as
\begin{equation}
\label{LaplaceWarp}
\tilde{R}_{4}-\tilde{\nabla}^2_6 e^{4\mathcal{A}}=-\frac{e^{2\mathcal{A}}}{2\text{Im}(\tau)}|G_3|^2-e^{-6\mathcal{A}}\left((\del \alpha)^2+(\del e^{4\mathcal{A}})^2\right)-\frac{1}{8\pi}e^{2\mathcal{A}}\left(T^m_m-T^{\mu}_{\mu}\right)^{\text{loc}}\, ,
\end{equation}
where $T_{MN}^{loc}=-\frac{2}{\sqrt{-g}}\frac{\delta S_{loc}}{\delta g^{MN}}$ is the energy momentum tensor of localized sources and $\tilde{R}_{4D}$ is the Ricci-scalar of the metric $\tilde{g}^4$.\par 
For the ansatz \eqref{C4ansatz}, the $F_5$ Bianchi identity can be expressed as
\begin{equation}
\label{LaplaceAlpha}
\tilde{\nabla}^2_6 \alpha=\frac{ie^{2\mathcal{A}}}{2\text{Im}(\tau)}G_3\cdot *_6 \bar{G_3}+2e^{-6\mathcal{A}}(\del \alpha)\cdot(\del e^{4\mathcal{A}})+\frac{T_3}{2\pi} e^{2\mathcal{A}}\rho_3^{\text{loc}}\, ,
\end{equation}
and by taking the difference between \eqref{LaplaceWarp} and \eqref{LaplaceAlpha} one obtains \cite{deAlwis:2003sn}
\begin{equation}
\label{GKPnogo}
\begin{split}
\tilde{\nabla}^2 \Phi^{-}=&\tilde{R}_{4D}+\frac{e^{2\mathcal{A}}}{\text{Im}(\tau)}|G_3^-|^2+e^{-6\mathcal{A}}|\del \Phi^-|^2+ e^{2\mathcal{A}}\frac{\Delta^{loc}}{2\pi}\, ,
\end{split}
\end{equation}
where $\tilde{R}_{4D}$ is the $4D$ Ricci scalar and we have defined
\begin{equation}
G_3^{\pm}\equiv\frac{1}{2}(*_6\pm i)G_3\, ,\quad \Phi^{\pm}\equiv e^{4\mathcal{A}}\pm\alpha \, , \quad \text{and}\quad \Delta^{loc}\equiv\frac{1}{4}\left(T^m_m-T^{\mu}_{\mu}\right)^{\text{loc}}- T_3\rho_3^{\text{loc}}\, .
\end{equation}
Integrating this expression over the compact manifold it follows that in the absence of localized sources that violate the BPS-like bound
\begin{equation}
\label{BPS-GKP}
\Delta^{loc}\geq 0\, ,
\end{equation}
the $4D$ vacuum energy $V\cdot M_{\rm P}^{-4}=\frac{1}{4}M_{\rm P}^{-2}\tilde{R}_{4D}$ is negative semi-definite,
\begin{equation}
\label{4dpotentialIIBapp}
V\cdot M_{\rm P}^{-4}=\int \frac{d^6 y\sqrt{g^6}}{16\pi\mathcal{V}_w\tilde{\mathcal{V}}_{w}} \left[-\,e^{8\mathcal{A}} \frac{\Delta}{2\pi}-|\del \Phi^{-}|^2\right]\leq 0\, ,
\end{equation}
where 
\begin{equation}
\begin{split}
\Delta&\equiv 2\pi\frac{|G_3^-|^2}{\text{Im}(\tau)}+\Delta^{loc}\, ,\quad  \tilde{\mathcal{V}}_{w}\equiv\int_{M_6}d^6y\sqrt{g^6}\, e^{6\mathcal{A}}\, ,\quad  \text{and}\quad \mathcal{V}_w\equiv\int_{M_6}d^6y\sqrt{g^6}\, e^{2\mathcal{A}}\, .
\end{split}
\end{equation}
In this notation the solutions of \cite{Giddings:2001yu} correspond to
\begin{equation}
G_3^-=\Delta^{loc}=\tilde{R}_{4D}=\Phi^{-}=0 \, .
\end{equation}
It is important to note that these solutions leave at least one modulus unfixed, the volume modulus. Clearly the inclusion of any further sources of positive energy cannot lead to a stable solution but rather lead to decompactification.\par 
This is true in particular for $\overline{D3}$ branes: the fact that they give a negative contribution to $\Delta^{loc}$ has the following simple interpretation: without K\"ahler moduli stabilization the presence of an $\overline{D3}$ brane leads to decompactification.

%%%%%%%%%%%%%%%%%%%%%%%%%%%%%%%%%%%%%%%%%%%%%%%%%%%%%%%%%%%%%%%%%%%%%%%%%%
\section{Contributions of fluxes and localized objects to the on-shell potential}\label{App:Contr.to.onshell.pot}
%%%%%%%%%%%%%%%%%%%%%%%%%%%%%%%%%%%%%%%%%%%%%%%%%%%%%%%%%%%%%%%%%%%%%%%%%%

In this appendix we would like to summarize the contributions of localized objects of tension $T_p$ and $p$-form fluxes to the on-shell potential \eqref{4dpotential_onshell} following \cite{Maldacena:2000mw}.
\subsection{The Contribution of $p$-form Fluxes}
\label{subsec:p-formfluxes}
For definiteness, we consider the case of a higher-dimensional $p$-form field strength, with action
\begin{equation}
S_{p-form}=\frac{M^{D-2}}{2}\int \left(-\frac{1}{2}F_p\wedge *F_p\right)\, .
\end{equation}
Then, the energy momentum tensor reads
\begin{equation}
T_{MN}=\frac{1}{2}\frac{1}{(p-1)!}F_{MO_2\cdots O_p}{F_N}^{O_2\cdots O_p}-\frac{1}{4}g_{MN}|F_p|^2\, .
\end{equation}
Let us now write $F_p=F_{p,ext}+F_{p,int}$, where $F_{p,ext}$ threads the $4$ non-compact directions if $p\geq 4$ and vanishes if $p<4$ while $F_{p,int}$ threads internal directions only. It follows that
\begin{equation}
T_P^P=\frac{2p-D}{4}\left(|F_{p,ext}|^2+|F_{p,int}|^2\right)\, \quad \text{and}\quad T_\mu^\mu=|F_{p,ext}|^2-|F_{p,int}|^2\, ,
\end{equation}
and the on-shell potential \eqref{4dpotential_onshell} is proportional to a (weighted) integral over the expression 
\begin{equation}
(D-6)T^{\mu}_{\mu}-4T^m_m=2(D-p-1)\underbrace{|F_{p,ext}|^2}_{\leq 0}+2(1-p)\underbrace{|F_{p,int}|^2}_{\geq 0}\, .
\end{equation}
It follows that internal components make a positive contribution only in the case $p=0$ and do not contribute when $p=1$ \cite{Maldacena:2000mw}. External components make a positive contribution for the top-form $p=D$ and a vanishing contribution for $p=D-1$. The latter two are of course equivalent to the former two by Hodge-duality.
\subsection{The Contribution of Localized Objects}
In addition, let us see how localized objects of spatial dimension $p$ and tension $T_p$ contribute. For this, consider a DBI-like action
\begin{equation}
S_{loc}=-T_p\int d^{p+1}\xi\, \sqrt{-\det\left(P[g]\right)}\, ,
\end{equation}
where $P[g]$ is the pullback of the ambient space metric on the localized objects world-volume. The energy momentum tensor is
\begin{equation}
T_{MN}=-\frac{T_p}{M^{D-2}}\Pi_{MN}(\Sigma)\cdot \delta(\Sigma)\, ,
\end{equation}
where $\Pi(\delta)$ is the projector on the $p-3$ cycle $\Sigma$ that the object wraps and $\delta(
\Sigma)$ is the transverse $\delta$ function. Hence, they contribute with
\begin{equation}
(D-6)T^{\mu}_{\mu}-4T^m_m=-4(D-p-3)\frac{T_p}{M^{D-2}}\delta(\Sigma)\, .
\end{equation}
For positive tension $T_p$ one immediately sees that positive contributions to the on-shell potential \eqref{4dpotential_onshell} come only from space time filling or co-dimension one objects. Also, negative tension objects of spatial dimension smaller than $D-3$ give a positive contribution, while co-dimension two objects do not contribute directly.

%%%%%%%%%%%%%%%%%%%%%%%%%%%%%%%%%%%%%%%%%%%%%%%%%%%%%%%%%%%%%%%%%%%%%%%%%%
\section{Gaugino condensates and the cosmological constant}\label{App:GauginoCondensateBackr.and.contr.to.onshell.cc}
%%%%%%%%%%%%%%%%%%%%%%%%%%%%%%%%%%%%%%%%%%%%%%%%%%%%%%%%%%%%%%%%%%%%%%%%%%

In this appendix we would like to provide the detailed derivation of the formulas used in section \ref{10D}. The perturbed $G_3$ profile has previously been calculated in \cite{Dymarsky:2010mf}. We provide also the contribution of all the terms under the integrand of \eqref{4dpotential_onshell}.\par
We are interested in the effects of the gaugino condensate in the bulk Calabi-Yau where (by definition) the back-reaction of fluxes on the geometry is negligible. Thus, in the following we work in the constant warp factor approximation where $R_{mn}= 0$. Hence, all results are valid to leading order in an inverse volume expansion.
\subsection{The $\bar{\lambda}\bar{\lambda}G_3$ coupling}\label{App:sub:lambdalambdacoupling}
We will follow appendix A of \cite{Dymarsky:2010mf} using results of appendix A of \cite{Lust:2008zd}. For vanishing worldvolume flux, the fermionic part of the $\kappa$-symmetric $D7$-brane action reads \cite{Marolf:2003ye,Marolf:2003vf,Martucci:2005rb}
\begin{equation}
\label{fermD7action}
S_{D7}^{ferm.}=i\pi\int_{\Sigma_8}e^{-\phi}\sqrt{-\det g}\bar{\theta}\left(1-\Gamma\right)\left(\Gamma^{\alpha}\mathcal{D}_{\alpha}-\frac{1}{2}\mathcal{O}\right)\theta\, ,
\end{equation}
where $\alpha,\beta...$ are $8d$ indices and $\theta$ is the two-component GS spinor. The operators $\mathcal{D}_{\alpha}$ and $\mathcal{O}$ are the pullbacks of the $10D$ operators
\begin{equation}
\label{SUSYgravitinoOp}
\mathcal{D}_{M}\theta=\nabla_{M}\theta+\frac{1}{4}\underline{H}_M\begin{pmatrix}
1 & 0 \\ 0 & -1
\end{pmatrix}\theta +\frac{1}{16}e^{\phi}\begin{pmatrix}
0 & \underline{F}\Gamma_M\\ -\sigma(\underline{F})\Gamma_M & 0
\end{pmatrix}\theta\, , 
\end{equation}
\begin{equation}
\label{SUSYdilatinoOp}
\mathcal{O}\theta=\underline{\del}\phi\theta +\frac{1}{2}\underline{H}\begin{pmatrix}
1 & 0 \\ 0 & -1
\end{pmatrix} \theta+ \frac{1}{16}e^{\phi}\begin{pmatrix}
0 & \Gamma^M \underline{F}\Gamma_M\\ -\Gamma^M \sigma(\underline{F})\Gamma_M & 0
\end{pmatrix}\theta\, ,
\end{equation}
where $\Gamma_M$ are $10D$ gamma matrices. Furthermore,
\begin{equation}
\Gamma=-\frac{i\sigma_2}{\sqrt{-g}}\frac{\epsilon_{\alpha_0...\alpha_7}}{8!}\Gamma^{\alpha_0...\alpha_7}\, ,
\end{equation}
and the map $\sigma$ reverses the order of indices. The barred spinor is defined as $\bar{\theta}=\theta^{\dagger}\Gamma^0$. Underlined tensors are contracted with gamma matrices. The action \eqref{fermD7action} is written in the (RR-)democratic formulation of type IIB supergravity. This means that $F=F_1+...+F_9$ and the equations of motion have to be supplemented by the duality constraint
\begin{equation}
\underline{F}\Gamma_*=\underline{F}\, ,
\end{equation}
in particular $\underline{F}_7\Gamma_*=\underline{F}_3$. Equivalently, one may work with $F_1,F_3$ and $F_5$ only, while doubling the contribution of $F_1$ and $F_3$ in \eqref{SUSYgravitinoOp} and \eqref{SUSYdilatinoOp}. In this case one only imposes the usual self-duality constraint $\underline{F}_5\Gamma_*=\underline{F}_5$. We will do so in the following.\par 
Imposing the $\kappa$-fixing gauge
\begin{equation}
\bar{\theta}\Gamma=-\bar{\theta}\, ,
\end{equation}
and keeping only three form fluxes with purely internal components the action can be written as
\begin{equation}
\begin{split}
S_{D7}^{ferm.}&=2\pi i\int d^8 \sigma\, e^{-\phi} \sqrt{-g}\left(\bar{\theta}\Gamma^{\alpha}(\nabla_{\alpha}-\frac{1}{2}\del_{\alpha}\phi)\theta\right.\\
&\left.+\bar{\theta}\left[\frac{1}{4}(\Gamma^{\alpha}\underline{H}_{\alpha}-\underline{H})\sigma_3+\frac{e^{\phi}}{8}(\Gamma^{\alpha}\underline{F}_3\Gamma_{\alpha}-\frac{1}{2}\Gamma_M \underline{F}_3\Gamma^M)\sigma_1\right]\theta\right)\, .
\end{split}
\end{equation}
This action is transformed to $10D$ Einstein frame using
\begin{equation}
g_{\alpha\beta}\longrightarrow e^{\phi/2}g_{\alpha\beta}\, ,\quad \Rightarrow\quad  \Gamma^{M}\longrightarrow e^{-\phi/4}\Gamma^M\, ,
\end{equation}
and the redefinition
\begin{equation}
\Theta\longrightarrow e^{-3\phi/8}\Theta\, .
\end{equation}
The result is
\begin{equation}
\begin{split}
S_{D7}^{ferm.}&=2\pi i\int d^8 \sigma\, \sqrt{-g}\bar{\theta}\Gamma^{\alpha}\nabla_{\alpha}\theta\\
&+e^{\phi/2}\bar{\theta}\left[\frac{e^{-\phi}}{4}(\Gamma^{\alpha}\underline{H}_{\alpha}-\underline{H})\sigma_3+\frac{1}{8}(\Gamma^{\alpha}\underline{F}_3\Gamma_{\alpha}-\frac{1}{2}\Gamma_M \underline{F}_3\Gamma^M)\sigma_1\right]\theta\, .
\end{split}
\end{equation}
Let us further massage this coupling of the schematic form $\bar{\theta}\theta G_3$. Two expressions need to be simplified,
\begin{equation}
\text{I:}\quad \Gamma^{\alpha}\underline{H}_{\alpha}-\underline{H}\, ,
\end{equation}
\begin{equation}
\text{II:}\quad \Gamma^{\alpha}\underline{F}_3\Gamma_{\alpha}-\frac{1}{2}\Gamma_M \underline{F}_3\Gamma^M\, .
\end{equation}
Let us decompose the three-form fluxes as follows:
\begin{equation}
\label{decompositionF}
F_3=F_3^{(0)}+F_3^{(1)}+F_3^{(2)}\, ,
\end{equation}
\begin{equation}
\label{decompositionH}
H_3=H_3^{(0)}+H_3^{(1)}+H_3^{(2)}\, ,
\end{equation}
where the upstairs index denotes the number of indices transverse to the divisor $\Sigma_4$ the $7$-brane wraps. By commuting $\Gamma$ matrices one can show that
\begin{equation}
\Gamma^{\alpha}\underline{F}\Gamma_{\alpha}-\frac{1}{2}\Gamma_M \underline{F}\Gamma^M=-2\frac{F_{ijk}}{3!}\left(P_{\perp}(\Gamma^i)\Gamma^j\Gamma^k+\Gamma^iP_{\perp}(\Gamma^j)\Gamma^k+\Gamma^i\Gamma^jP_{\perp}(\Gamma^k)\right)\, .
\end{equation}
Here, $P_{\perp}(\Gamma^i)=\Gamma^i$ if the index $i$ is transverse to $\Sigma_4$ while it vanishes otherwise. Under the decomposition \eqref{decompositionF} this reads
\begin{equation}
\Gamma^{\alpha}\underline{F}\Gamma_{\alpha}-\frac{1}{2}\Gamma_M \underline{F}\Gamma^M=-2\underline{F}^{(1)}-4\underline{F}^{(2)}\, .
\end{equation}
Similarly, one shows that
\begin{equation}
\Gamma^{\alpha}\underline{H}_{\alpha}-\underline{H}=2\underline{H}-3\frac{H_{ijk}}{3!}P_{\perp}(\Gamma^i)\Gamma^j\Gamma^k=2\underline{H}^{(0)}+\underline{H}^{(1)}\, .
\end{equation}
Thus, the $\bar{\theta}\theta G_3$ coupling reads
\begin{equation}
S_{D7}\supseteq \frac{\pi}{2} i \int d^8 \sigma\, \sqrt{-g}e^{\phi/2}\bar{\theta}\left(e^{-\phi}(2\underline{H}^{(0)}+\underline{H}^{(1)})\sigma_3-(\underline{F}^{(1)}+2\underline{F}^{(2)})\sigma_1\right)
 \theta\, .
\end{equation}
To perform the dimensional reduction one decomposes the $10D$ $\Gamma$-matrices as
\begin{equation}
\Gamma^{\mu}=e^{-\mathcal{A}}\tilde{\gamma}^{\mu}\otimes \id\, ,\quad \mu=0,...,3\, ,
\end{equation}
\begin{equation}
\Gamma^i=\tilde{\gamma}_{*}\otimes \gamma_{6D}^{i-3}\, ,\quad i=4,...,9
\end{equation}
with $4D$ respectively $6D$ gamma matrices $\tilde{\gamma}^{\mu}$ and $\gamma^{i-3}_{6D}$.\par 
The dimensional reduction ansatz for the gaugino is
\begin{equation}
\theta_1=\frac{1}{4\pi} e^{-2\mathcal{A}}\lambda_D \otimes \eta_1+ c.c. \, , \quad \theta_2=-\frac{1}{4\pi} e^{-2\mathcal{A}}\lambda_D \otimes \eta_2+ c.c.
\end{equation}
Here, $\lambda_D$ is a $4D$ Dirac spinor of positive chirality, and $c.c.$ denotes charge conjugation. Moreover, $\eta_1$ and $\eta_2$ are the $6D$ Weyl spinors that appear in the $10D$ SUSY parameters:
\begin{equation}
\epsilon_1=\xi\otimes \eta_1 +c.c.\, ,\quad \epsilon_2=\xi\otimes \eta_2 +c.c.
\end{equation}
For $O3/O7$ orientifolds with ISD fluxes and $D3/D7$-branes, the relation between $\eta\equiv\eta_1$ and $\eta_2$ is \cite{Grana:2004bg}
\begin{equation}
\eta_2=i \eta_1\, .
\end{equation} 
Thus, we may write
\begin{equation}
\theta_1=\psi+\psi^c\, , \quad \theta_2=-i(\psi-\psi^c)\, , \quad \psi\equiv \frac{1}{4\pi}e^{-2\mathcal{A}}\lambda_D\otimes \eta\, ,
\end{equation}
where the superscript $^c$ denotes charge conjugation.\par
With this one calculates that
\begin{equation}
e^{4\mathcal{A}}\bar{\theta} \Gamma_{ijk}\sigma_3\theta=2e^{4\mathcal{A}}\overline{\psi^c}\Gamma_{ijk}\psi-c.c.=-\frac{i}{8\pi^2}\overline{\lambda_D^c}\lambda_D(\eta^c)^{\dagger} \gamma_{ijk}\eta-c.c.\, ,
\end{equation}
and
\begin{equation}
e^{4\mathcal{A}}\bar{\theta}\Gamma_{ijk}\sigma_1\theta=-2ie^{4\mathcal{A}} \overline{\psi^c}\Gamma_{ijk}\psi-c.c.=-\frac{1}{8\pi^2}\overline{\lambda_D^c}\lambda_D(\eta^c)^{\dagger} \gamma_{ijk}\eta-c.c.
\end{equation}
Here, the $4D$ barred spinor is the usual Dirac conjugate $\overline{\lambda_D}=i\lambda_D^{\dagger}\tilde{\gamma}^0$.\par 
Let us define
\begin{equation}
G_3^{(i,j)}:=F_3^{(i)}-ie^{-\phi}H_3^{(j)}\, ,
\end{equation}
\begin{equation}
\mathcal{G}_3:=G_3^{(1,1)}+2G_3^{(2,0)}\, .
\end{equation} 
Then we may write the result as
\begin{equation}
\label{finalG3Ocoupling}
\begin{split}
S_{D7}&\supseteq
\int_{M_{10}}d^{10}x \, \sqrt{-g} \mathcal{L}^{loc}_{\lambda\lambda}\equiv \int_{M_4\times \Sigma_4} d^8 \sigma \sqrt{-g}\, e^{-4\mathcal{A}}e^{\phi/2}\frac{i\overline{\lambda_D^c}\lambda_D}{16\pi} \,\Omega\cdot \mathcal{G}_3+c.c.\\
&=\int_{M_4\times \Sigma_4} d^8 \sigma \sqrt{-g}e^{-4\mathcal{A}}e^{\phi/2}\, \frac{\bar{\lambda}_{\dot{\alpha}}\bar{\lambda}^{\dot{\alpha}}}{16\pi} \,\Omega\cdot \mathcal{G}_3+c.c.=\int_{M_{10}} \delta^{(0)}_De^{-4\mathcal{A}}e^{\phi/2}\frac{\bar{\lambda}_{\dot{\alpha}}\bar{\lambda}^{\dot{\alpha}}}{16\pi}\mathcal{G}_3\wedge *\Omega +c.c.\, .
\end{split}
\end{equation}
where we have also used that the holomorphic three-form $\Omega$ is given in terms of the $6D$ spinor $\eta$ as
\begin{equation}
\label{holomorphic3form}
\Omega_{ijk}=(\eta^c)^{\dagger}\gamma_{ijk}\eta\, ,
\end{equation}
and we have written the action as a $10D$ integral by introducing the scalar delta-function $\delta_D^{(0)}$ that localizes on the divisor $\Sigma_4$. It satisfies \cite{Baumann:2010sx}
\begin{equation}
2\pi\delta^{(0)}_D= \nabla^2 \text{Re} \log h=2g^{i\bar{i}}\nabla_i\nabla_{\bar{i}}\text{Re}\log h\, ,
\end{equation}
where the holomorphic function $h$ defines the divisor that the seven-brane wraps via $h=0$.\par
Moreover, in the second line of \eqref{finalG3Ocoupling} we have chosen the $4D$ Weyl-representation where the spinor $\lambda_D$ takes the form $(0,\bar{\lambda}^{\dot{\alpha}})^T$. Finally, we used that $\overline{\lambda_D^c}\lambda_D=\lambda_D^TC\lambda_D=-i \bar{\lambda}_{\dot{\alpha}}\bar{\lambda}^{\dot{\alpha}}$. From now on we will write $\bar{\lambda}\bar{\lambda}=\bar{\lambda}_{\dot{\alpha}}\bar{\lambda}^{\dot{\alpha}}$ and $\lambda\lambda= (\bar{\lambda}\bar{\lambda})^*$.\par
One further simplification can be made: $\Omega\cdot \mathcal{G}_3$ projects on the $(0,3)$ piece of $\mathcal{G}_3$. Because the submanifold that the $7$-branes wrap is complex \cite{federer1969geometric,Harvey:1982xk} (see also \cite{Koerber:2005qi} for flux compactifications beyond GKP) the components of $F_3$ and $H_3$ that have precisely zero or two legs transverse to the brane are of Hodge type $(2,1)\oplus (1,2)$ which implies that $\mathcal{G}_3\cdot \Omega=G_3\cdot \Omega$. Thus, we finally arrive at
\begin{equation}
S_{D7}\supseteq\int_{M_{10}} \pi \delta^{(0)}_De^{\phi/2}e^{-4\mathcal{A}}\frac{\bar{\lambda} \bar{\lambda}}{16\pi^2}\,G_3\wedge *\Omega +c.c.\, .
\end{equation}
\subsection{The $G_3$ equations of motion and their solution}\label{App:sub:G3eom.and.sol}
In the presence of a non-vanishing expectation value $\langle \lambda \lambda \rangle$, the equations of motion and Bianchi identity of $G_3$ read
\begin{equation}
\label{G3eom}
d\Lambda -\frac{d\tau}{\tau-\bar{\tau}}\wedge (\Lambda+ \bar{\Lambda})=dX-\frac{d\tau}{\tau-\bar{\tau}}\wedge (X+\bar{X})\, ,
\end{equation}
\begin{equation}
\label{G3Bianchi}
dG^+-\frac{d\tau}{\tau-\bar{\tau}}\wedge(G^++\bar{G^+})=dG^--\frac{d\tau}{\tau-\bar{\tau}}\wedge(G^-+\bar{G^-})\, ,
\end{equation}
where $\Lambda\equiv e^{4\mathcal{A}}*_6 G_3-i\alpha G_3$ and $X$ is defined via
\begin{equation}
d^4x \wedge \left(dX-\frac{d\tau}{\tau-\bar{\tau}}\wedge (X+\bar{X})\right)=-\frac{i}{2\pi}\left(\tau d\left(\frac{\del \mathcal{L}^{loc}_{\lambda\lambda}}{\del dC_2}\right)+d\left(\frac{\del \mathcal{L}^{loc}_{\lambda\lambda}}{\del dB_2}\right)\right)\, ,
\end{equation}
which determines
\begin{equation}
X=e^{-\phi/2}\frac{\langle \lambda\lambda \rangle}{16\pi^2}\delta_D *_6 \bar{\Omega}\, .
\end{equation} The ISD solutions correspond to $\Lambda=e^{4\mathcal{A}}G^-=0$.\par 
We expand the above equation to linear order in $\langle \lambda\lambda \rangle$. Equation \eqref{GKPnogo} implies that to this order $\Phi^-=0$. Therefore\footnote{We thank the authors of \cite{Gautason:2018toappear} for pointing out an erroneous factor of two that we have corrected in what follows.},
\begin{equation}
\Lambda=\Phi^+ G_3^-+\Phi^- G_3^+=2e^{4\mathcal{A}}G_3^-+\mathcal{O}(|\langle \lambda\lambda \rangle|^2)\, ,
\end{equation}
and hence
\begin{equation}
d(e^{4\mathcal{A}}G_3^-)=\frac{1}{2}dX\, .
\end{equation}
Note that it is not possible to simply set $e^{4\mathcal{A}}G_3=\frac{1}{2}X$ because by definition $G_3^-$ is IASD while $X$ is ISD. In the flat Calabi-Yau limit $R_{ij}=0$ the solution was derived in \cite{Baumann:2010sx}. It is shown that bulk IASD fluxes of Hodge type $(1,2)$ are sourced:
\begin{equation}
\label{bulkIASD}
\begin{split}
&(e^{4\mathcal{A}}G_3^-)_{i\bar{j}\bar{k}}=-\frac{i}{2\pi} e^{-\phi/2}\frac{\langle \lambda\lambda \rangle}{16\pi^2}(\nabla_i \nabla_l \text{Re}(\log h))g^{l\bar{m}}\bar{\Omega}_{\bar{m}\bar{j}\bar{k}}\, ,\\
\Rightarrow\quad &2\pi e^{\phi}|G_3^-|^2=\frac{8}{\pi}e^{-8\mathcal{A}}\left|\frac{\langle \lambda\lambda \rangle}{16\pi^2}\right|^2 |\nabla_i\nabla_j \text{Re}\log h|^2\, ,
\end{split}
\end{equation}
using $|\Omega|^2\equiv \frac{1}{3!}\Omega_{ijk}\bar{\Omega}^{ijk}=8$.
Eq. \eqref{G3eom} only determines the IASD flux. The ISD part of $G_3$ follows from imposing the Bianchi-identity \eqref{G3Bianchi}, i.e.
\begin{equation}
dG_3^+=dG_3^-= \frac{1}{2}e^{-4\mathcal{A}}dX\, ,
\end{equation}
which is solved by
\begin{equation}
\label{ISDG3}
e^{4\mathcal{A}}G_3^+=\frac{1}{2}X=\frac{1}{2\pi}e^{-\phi/2}\frac{\langle \lambda\lambda \rangle}{16\pi^2}(g^{i\bar{j}}\nabla_i \nabla_{\bar{j}} \text{Re}(\log h)) \,\overline{\Omega}\, .
\end{equation}
This localized ISD flux was computed in \cite{Dymarsky:2010mf}.
\subsection{The contributions to the on-shell potential}\label{App:sub:lambdalambda.contr.to.cc}
We are now ready to derive the contribution of the fermionic action \eqref{fermD7action} to $\Delta_{loc}$. For this one has to vary the action with respect to the vielbein $e^{M}_a$ defined by
\begin{equation}
g^{MN}=e^M_a e^N_b \eta^{ab}\, .
\end{equation}
There are two contributions: the first is due to the variation of the volume form. Because the $D7$ has co-dimension two, this will drop out in $\Delta_{loc}$. The second contribution comes from varying the curved space $\Gamma$-matrices $\Gamma^M=e^M_a \Gamma^a$ with constant matrices $\Gamma^a$. Then, by using $g^{mn}\frac{\del}{\del g^{mn}}=\frac{1}{2}e^m_c\frac{\del}{\del e^m_c}$ it is straightforward to obtain the desired negative contribution to $\Delta_{loc}$,
\begin{equation}
\begin{split}
\Delta_{loc}&\supseteq \Delta_{loc}^{\lambda\lambda}\equiv -\frac{1}{4}e^m_c\frac{\del\mathcal{L}^{loc}_{\lambda\lambda}}{\del e^m_c}=-\frac{3\pi}{8} e^{\phi/2}e^{-4\mathcal{A}}\frac{\langle \bar{\lambda}\bar{\lambda} \rangle}{16\pi^2} \,\Omega\cdot G_3\, \delta^{(0)}_D+c.c.\\
&\sim  -e^{-8\mathcal{A}}\left|\frac{\langle \lambda\lambda \rangle}{16\pi^2}\right|^2 |g^{i\bar{j}}\nabla_i \nabla_{\bar{j}} \text{Re}(\log h)|^2<0\, .
\end{split}
\end{equation}
Bulk IASD fluxes \eqref{bulkIASD} contribute to $\Delta$ with
\begin{equation}
\Delta_{bulk}^{\lambda\lambda}\sim  e^{-8\mathcal{A}}\left|\frac{\langle \lambda\lambda \rangle}{16\pi^2}\right|^2 |\nabla_i\nabla_j \text{Re} \log h|^2>0\, .
\end{equation}
Thus finally,
\begin{equation}\label{eq:DeltaLambdaLambda}
\Delta\supseteq \Delta^{\lambda\lambda}=  e^{-8\mathcal{A}}\left|\frac{\langle \lambda\lambda \rangle}{16\pi^2}\right|^2 \left[\alpha|\nabla_i\nabla_j \text{Re} \log h|^2-\beta|g^{i\bar{i}}\nabla_i\nabla_{\bar{i}} \text{Re} \log h|^2\right]\, .
\end{equation}
Here, we have introduced two undetermined positive $\mathcal{O}(1)$ factors $\alpha,\beta$ for the following reason\footnote{In an earlier version of this paper the "smoothing regularization" was implemented too literally, i.e. specified numerical values of the $\alpha,\beta$ were given with $\alpha/\beta=4/3$. This is problematic for two reasons: First, the regularization scheme is simply too ad-hoc for such a stringent prediction. Second, due to an erroneous factor of two that was pointed out to us by the authors of \cite{Gautason:2018toappear}, one would actually have $\alpha/\beta=2/3$ which would be inconsistent with the existence of supersymmetric AdS vacua of KKLT type. We thank the authors of \cite{Gautason:2018toappear} for discussions about this point.}. The integral of both terms in eq. \eqref{eq:DeltaLambdaLambda} over the internal manifold diverges toward the position of the seven-brane stacks, and has to be regularized. We expect generically that cutting off the integrals at the string-scale (or smoothing $\text{Re}\log h$ over an order $l_s$ interval transverse to the seven-branes) gives the correct answer up to uncertainties of order unity. We parametrize this regulator dependence by the $\mathcal{O}(1)$ factors $\alpha,\beta$.

For a set of $n$ stacks of seven-branes that wrap divisors $\Sigma_a$, $a=1,...,n$, the contribution to $\Delta$ becomes
\begin{equation}
\Delta^{\lambda\lambda}=e^{-8\mathcal{A}} \left[\alpha \left|\sum_{a=1}^{n}\frac{\langle \lambda\lambda \rangle_a}{16\pi^2}\nabla_i\nabla_j \text{Re} \log h_a\right|^2-\beta\sum_{a=1}^{n}\left|\frac{\langle \lambda\lambda \rangle_a}{16\pi^2} g^{i\bar{i}}\nabla_{i}\nabla_{\bar{i}} \text{Re} \log h_a\right|^2\right]\, ,
\end{equation}
where upon integration regularization is implicit as explained above.

Because the contribution to the cosmological constant is determined by integrating this expression over the internal manifold we may partially integrate
\begin{equation}
\begin{split}
&\int_{M_6} *_6\,|\nabla_i\nabla_j g|^2=\int_{M_6} *_6\,g^{i\bar{i}}g^{j\bar{j}}\nabla_i \nabla_{j}g\nabla_{\bar{i}}\nabla_{\bar{j}}g\overset{p.I.}{=}-\int_{M_6} *_6\,(g^{j\bar{j}}(g^{i\bar{i}}\nabla_i\nabla_{\bar{i}}\nabla_jg) \nabla_{\bar{j}}g)\\
&=-\int_{M_6} *_6\,(g^{j\bar{j}}g^{i\bar{i}}(\nabla_j \nabla_i\nabla_{\bar{i}} g+\frac{1}{2} \underbrace{[\nabla^2,\nabla_j]g}_{=R_{jk}\del^k g=0})\nabla_{\bar{j}}g)\overset{p.I.}{=}\int_{M_6} *_6\,|g^{i\bar{i}}\nabla_i\nabla_{\bar{i}}g|^2\, .
\end{split}
\end{equation}
for any real function $g$.\par 
Therefore, we may write
\begin{equation}
\label{finaldeltalambda}
\Delta^{\lambda\lambda}=e^{-8\mathcal{A}} \left[\alpha\left|\sum_{a=1}^{n}\frac{\langle \lambda\lambda \rangle_a}{16\pi^2}\nabla_i\nabla_j\text{Re} \log h_a\right|^2-\beta\sum_{a=1}^{n}\left|\frac{\langle \lambda\lambda \rangle_a}{16\pi^2}\nabla_i\nabla_j \text{Re} \log h_a\right|^2\right]\, .
\end{equation}
Crucially, for a single gaugino condensate
\begin{equation}
\Delta^{\lambda\lambda}=(\alpha-\beta)e^{-8\mathcal{A}}\left|\frac{\langle \lambda\lambda \rangle}{16\pi^2}\nabla_i\nabla_j \text{Re} \log h\right|^2\, ,
\end{equation}
which implies that if gaugino condensation is the only ISD-breaking source, the $4D$ potential energy is given by
\begin{equation}
V\cdot M_{\rm P}^{-4}=-\frac{\alpha-\beta}{32\pi^2\mathcal{V}_w\tilde{\mathcal{V}}_w}\int d^6 y \sqrt{g}\left|\frac{\langle \lambda\lambda \rangle}{16\pi^2}\nabla_i\nabla_j \text{Re} \log h\right|^2\, .
\end{equation}
Demanding the existence of supersymmetric KKLT AdS-vacua implies that $\alpha-\beta>0$.

For two condensates the first term in \eqref{finaldeltalambda} can be made small by giving the condensates opposite phases. In this case the two terms in \eqref{finaldeltalambda} cancel each other at least partially. Of course, for a SUSY vacuum the final result can never be positive but we find it very conceivable that a small SUSY breaking source (like an $\overline{\text{D3}}$ brane) can lift such a vacuum to de Sitter. 

%%%%%%%%%%%%%%%%%%%%%%%%%%%%%%%%%%%%%%%%%%%%%%%%%%%%%%%%%%%%%%%%%%%%%%%%%%
\section{A volume suppressed contribution }\label{App:vol.supp.contr.softmasses}
%%%%%%%%%%%%%%%%%%%%%%%%%%%%%%%%%%%%%%%%%%%%%%%%%%%%%%%%%%%%%%%%%%%%%%%%%%

Here we would like to address an example of a volume suppressed contribution to the on-shell potential \eqref{4dpotentialIIB} that we must neglect for consistency of our expansion scheme. Before uplifting the flux configuration is as follows
\begin{equation}
G_3=\underbrace{G_3^{(2,1)}}_{\text{harmonic}}+\underbrace{G_3^{\lambda\lambda}}_{\text{particular solution}}\, ,
\end{equation}
where $G_3^{\lambda\lambda}$ is the profile that has been computed in \cite{Baumann:2010sx,Dymarsky:2010mf}. Therefore, the localized $(0,3)$ piece in $G_3^{\lambda\lambda}$ is in the same cohomology class as the harmonic $G_3$ piece of Hodge type $(0,3)$ that would have existed without non-perturbative stabilization. The value of the gaugino condensate is tied to the profile $G_3^{\lambda\lambda}$ via the equations of motion. Thus, the condensate will dynamically find its correct value such that the localized $(0,3)$ piece in $G_3^{\lambda\lambda}$ can account for all of $W_0=\int G_3\wedge \Omega$.\par
We start with this supersymmetric situation and deform it by including a SUSY breaking source at the bottom of a warped throat. Such a source will push the value of the volume modulus towards larger values and thereby reduce the magnitude of the condensate. 
\begin{equation}
\langle \lambda\lambda\rangle_0\longrightarrow \langle \lambda\lambda\rangle_1\, .
\end{equation}
Because the field profile $G_3^{\lambda\lambda}$ is still tied to the value of the condensate, in order for this to be consistent with flux quantization and a negligible pull on the complex structure moduli there must develop a harmonic $(0,3)$ piece in $G_3$,
\begin{equation}
G_3\longrightarrow G_3+g^{(0,3)}\overline{\Omega}\, ,
\end{equation}
such that $\int G_3\wedge \frac{\Omega}{||\Omega||}$ remains constant. Using the identities
\begin{equation}
||\Omega||^2=\int_{M_6}\Omega\wedge \overline{\Omega}=8\mathcal{V}\,\quad \text{and}\quad \int_{M_6}\delta_D\Omega\wedge \overline{\Omega}=8\text{Vol}(\Sigma_4)\, , 
\end{equation}
this fixes $g^{(0,3)}$ to be
\begin{equation}
g^{(0,3)}=e^{-4\mathcal{A}}|_{\Sigma}e^{-\phi/2}\frac{\text{Vol}(\Sigma)_1}{\mathcal{V}_1}\left(\frac{\langle \lambda\lambda\rangle_0}{16\pi^2} \frac{\text{Vol}(\Sigma)_0}{\text{Vol}(\Sigma)_1}\sqrt{\frac{\mathcal{V}_1}{\mathcal{V}_0}} -\frac{\langle \lambda\lambda\rangle_1}{16\pi^2}\right)\, .
\end{equation}
Here the subindices $0$ and $1$ denote pre- respectively post uplift situations. Clearly this constant flux piece contributes to $\Delta^{loc}$ and therefore contributes a term in the on-shell potential \eqref{4dpotential_onshell}. However, to leading order in the inverse volume expansion this term must be neglected. This is clear intuitively because only the value of $G_3^{(0,3)}$ at the position of the $7$-brane divisor $\Sigma_4$ enters the on-shell potential \eqref{4dpotentialIIB}. If part of the localized term is traded for a constant term to leading order in the inverse transverse volume only a reduction in the localized $G_3^{(0,3)}$ component is seen. Let us show this explicitly: the constant flux contributes a term to the on-shell potential
\begin{equation}\label{eq:VfluxLocalAppendix}
V_{g^{(0,3)}}\cdot M_{\rm P}^{-4}\equiv \frac{3}{16\pi}\frac{\text{Vol}(\Sigma)_1^2}{\mathcal{V}_1^3}\frac{\langle \bar{\lambda}\bar{\lambda}\rangle_1}{16\pi^2}\left[ \frac{\langle \lambda\lambda\rangle_0}{16\pi^2} \frac{\text{Vol}(\Sigma)_0}{\text{Vol}(\Sigma)_1}\sqrt{\frac{\mathcal{V}_1}{\mathcal{V}_0}} -\frac{\langle \lambda\lambda\rangle_1}{16\pi^2} \right]+c.c.
\end{equation}
Because all volumes are pushed towards larger values and the value of the condensate decreases exponentially with the volume we have $\langle \lambda\lambda\rangle_1<\langle \lambda\lambda\rangle_0$. Hence this contribution is positive.\par
If a warped tension $\delta V/M_{\rm P}^4=\frac{e^{4\mathcal{A}_0}T_3}{(4\pi)^2\mathcal{V}^2}$ is added to the setup it is clear that if the warp factor $e^{\mathcal{A}_0}$ is sufficiently small, the change in potential is given precisely by this term. Using this information and our knowledge about the off-shell potential we can deduce how far the volume modulus shifts. We assume the usual exponential dependence
\begin{equation}
|\langle \lambda\lambda\rangle| \sim e^{-a\text{Vol}(\Sigma)}\, ,
\end{equation}
and choose to calculate to leading order in $1/\text{Vol}(\Sigma)$. In this approximation we need only consider the shift of $\langle \lambda\lambda\rangle$ itself and not of explicit volume powers that appear in the above formulas. The shift in the potential in equation eq.~\eqref{PotentialSingleCondensate} induced by a small shift $\delta \text{Vol}(\Sigma)$ is
\begin{equation}
\delta V\cdot M_{\rm P}^{-4}\sim a\, \delta \text{Vol}(\Sigma)\cdot |V_0\cdot M_{\rm P}^{-4}|\sim a\, \delta \text{Vol}(\Sigma)\cdot \frac{\text{Vol}(\Sigma)}{\mathcal{V}^2}\left|\langle \lambda\lambda \rangle\right|^2\, ,
\end{equation}
while
\begin{equation}
V_{g^{(0,3)}}\cdot M_{\rm P}^{-4}\sim a\, \delta \text{Vol}(\Sigma)\cdot \frac{\text{Vol}(\Sigma)^2}{\mathcal{V}^3}\left|\langle \lambda\lambda \rangle\right|^2\, .
\end{equation}
Hence the positive contribution coming from $g^{(0,3)}$ is suppressed by the transverse volume and must be neglected. The shift in the volume modulus is related to the warped tension via
\begin{equation}
a\, \delta \text{Vol}(\Sigma)\sim \frac{e^{4\mathcal{A}_0} T_3}{\text{Vol}(\Sigma)|\langle \lambda\lambda \rangle|^2}\, ,
\end{equation}
which determines the value of $g^{(0,3)}$ to be
\begin{equation}
g^{(0,3)}\sim e^{-\phi/2}\frac{e^{4\mathcal{A}_0} T_3}{\mathcal{V}}\, .
\end{equation}
Therefore, similarly to the situation of non-supersymmetric classical ISD solutions the value $g^{(0,3)}$ parametrizes SUSY breaking for instance by inducing gaugino mass-terms on $D7$ and $D3$ branes \cite{Camara:2003ku,Grana:2003ek,Camara:2004jj}.
%%%%%%%%%%%%%%%%%%%%%%%%%%%%%%%%%%%%%%%%%%%%%%%%%%%%%%%%%%%%%%%%%%%%%%%%%%
\section{Radial profiles of throat perturbations}\label{App:RadialProfiles}
%%%%%%%%%%%%%%%%%%%%%%%%%%%%%%%%%%%%%%%%%%%%%%%%%%%%%%%%%%%%%%%%%%%%%%%%%%
Our analysis relies on the fact that perturbations that are sourced at the bottom of a warped throat fall-off towards the bulk Calabi-Yau fast enough to have no significant effect on the integrand of the on-shell potential \eqref{4dpotentialIIB}. Therefore, in this appendix we provide details about the radial scaling of throat modes. For our purpose it is sufficient to determine the radial scaling of IR perturbations.
\subsection{Throat perturbations sourced in the IR}
In order to be able to determine how \eqref{4dpotentialIIB} is affected by a SUSY breaking source at the bottom of a warped throat, we need to know the approximate radial dependence of field profiles $\phi(r)$ that correspond to perturbations of the KS throat that are sourced in the IR. We do not need to know their field profiles in full detail but only need to estimate the ratio $\phi(r_{UV})/\phi(r_{IR})$ in terms of powers of the warp factor in the IR $e^{\mathcal{A}_0}\equiv \left.e^{\mathcal{A}}\right|_{IR}$ and the $\rho$-modulus mass $m_{\rho}$. This makes the analysis much easier as for this purpose one may approximate parts of the throat by $AdS_5\times T^{1,1}$. In this approximation the field profiles can be determined analytically.\par   
In more detail: away from the tip of the throat the geometry is well approximated by the Klebanov-Tseytlin (KT) solution \cite{Klebanov:2000nc}
\begin{equation}
ds^2=e^{2\mathcal{A}(r)}dx^2+e^{-2\mathcal{A}}\left(dr^2+r^2ds^2_{T^{1,1}}\right)\, .
\end{equation}
with
\begin{equation}
e^{-4\mathcal{A}}=L^4\frac{\ln(r/r_s)}{r^4}\, .
\end{equation}
The KT solution is valid for $r\gg r_s$. For our purposes the logarithmic running will be irrelevant and we approximate the throat by $AdS_5\times T^{1,1}$, i.e. 
\begin{equation}
e^{-4\mathcal{A}}\longrightarrow L^4/r^4\, .
\end{equation}
For this case the analysis of radial profiles of throat perturbation was done for instance in \cite{Gandhi:2011id}. 
\subsection{Scalar perturbations}\label{App:sub:RadialProfiles.scalar}
Since we are interested in perturbations that are sourced only deep in the IR, only the homogeneous modes of perturbations need to be studied. For scalar perturbations these are solutions to Laplace' equation
\begin{equation}
\nabla^2 \phi=0\, .
\end{equation}
This is solved by a product ansatz $\phi(r,\Psi)=\sum_{i}\phi_i(r)Y_i(\Psi)$, where the functions $Y_i(\Psi)$ are eigenfunctions of the Laplacian of $T^{1,1}$
\begin{equation}
\nabla_{T^{1,1}}^2Y_i=-\lambda_i^2 Y_i\, .
\end{equation}
Because the $10D$ Laplacian reads $\nabla^2=e^{2\mathcal{A}}\left[\del_r^2+\frac{5}{r}\del_r+\frac{1}{r^2}\nabla_{T^{1,1}}^2\right]$, the remaining radial equation of motion reads
\begin{equation}
\label{radialeom}
\del_r^2\phi_i+\frac{5}{r}\del_r\phi_i-\frac{\lambda_i^2}{r^2}\phi_i=0\, ,
\end{equation}
and is solved by 
\begin{equation}
\phi_i(r)=\sum_{i}a_i\, r^{\Delta_i-4}+b_i\, r^{-\Delta_i}\, ,\quad \Delta_i \equiv 2+\sqrt{4+\lambda_i^2}\, .
\end{equation}
For an infinite throat perturbations that are sourced in the IR are the normalizable ones. Because $\Delta_i\geq 4$, the normalizable mode corresponds to $\phi \propto r^{-\Delta_i}$. The KK zero mode of $T^{1,1}$ (i.e. the mode with $\lambda_0=0$) falls off least rapidly towards the UV, i.e.
\begin{equation}
\phi_0(r_{UV})/\phi_0(r_{IR})\sim \left(\frac{r_{UV}}{r_{IR}}\right)^{-4}= e^{4\mathcal{A}_0}\, .
\end{equation}
In the following we would like to see how this behavior is changed when the throat is finite. In this case the would-be non-normalizable mode is not frozen and can in principle play a substantial role as well. This is particularly clear for the KK-zero mode: if it is sourced in the IR it will simply adjust its VEV along the whole throat. In this case its radial profile is determined entirely by the non-normalizable mode.\par
To discuss this we have to know the UV boundary conditions. Although we do not know these precisely we can encode them in a 'UV-potential' in the effective $5d$ theory. First we would like to discuss the KK-modes with $\lambda_i\neq 0$. For now we set the UV-potential to zero, i.e. impose $\del_r\phi_i|_{r=r_{UV}}=0$. In this case the solution to the equations of motion can be expressed in terms of the value $\phi_{IR}\equiv \phi(r_{IR})$ as follows
\begin{equation}
\phi(r)=\frac{(\Delta-4)e^{(4-2\Delta)\mathcal{A}_0}r^{-\Delta}+\Delta r^{\Delta-4}}{(\Delta-4)e^{(4-2\Delta)\mathcal{A}_0}+\Delta}\phi_{IR}\, ,
\end{equation}
which implies
\begin{equation}
\phi_{UV}/\phi_{IR}=\frac{(2\Delta-4) e^{\Delta\mathcal{A}_0}}{e^{(2\Delta-4)\mathcal{A}_0}\Delta+(\Delta-4)}=
\begin{cases}
2 e^{\Delta\mathcal{A}_0}& \text{for}\quad 1\ll \lambda^2\, ,\\
\sim e^{\Delta\mathcal{A}_0}\, , & \text{for}\quad  \lambda\sim 1\, ,\\
\frac{4e^{4\mathcal{A}_0}}{4e^{4\mathcal{A}_0}+\frac{\lambda^2}{4}}& \text{for}\quad \lambda^2\ll 1\, .\\
\end{cases}\, 
\end{equation}
Thus, up to factors of order unity the dependence on the warp factor $e^{\mathcal{A}_0}$ can be correctly determined by retaining the normalizable mode only as long as $\lambda^2\gg e^{4\mathcal{A}_0}$. For all but the zero mode this is the case.
Clearly, including a UV-potential can only further reduce $\phi_{UV}/\phi_{IR}$.\par 
Next we consider those fields that have no bulk potential (i.e. KK-zero modes of $T^{1,1}$) but are stabilized by global effects of the bulk CY, like the axio-dilaton in generic flux compactifications. For these fields the UV potential determines the field profile entirely so we cannot neglect it. We approximate the UV-potential around its minimum by
\begin{equation}
V_{UV}(\phi)=\frac{m_{UV}^2}{2k}\phi^2\, .
\end{equation}
Now the UV-boundary condition reads $r\del_r \phi|_{r=r_{UV}}=-\frac{m_{UV}^2}{k^2}\phi(r_{UV})$ and the boundary values $\phi_{UV}$ and $\phi_{IR}$ are related by
\begin{equation}
\phi_{UV}/\phi_{IR}=\frac{4k^2e^{4\mathcal{A}_0}}{m_{UV}^2(1-e^{4\mathcal{A}_0})}\left(1+\frac{4k^2e^{4\mathcal{A}_0}  }{m_{UV}^2(1-e^{4\mathcal{A}_0})}\right)^{-1}
\end{equation}
Again, for stabilization $m_{UV}^2/k^2\gg e^{4\mathcal{A}_0}$ one obtains
\begin{equation}
\phi_{UV}/\phi_{IR}\approx \frac{4k^2e^{4\mathcal{A}_0}}{m_{UV}^2}\, .
\end{equation}
For the UV-potential we may not assume that $m_{UV}\sim k$. Thus, by only keeping the normalizable mode one would miss a factor of $k^2/m_{UV}^2$. In the setting of type IIB flux compactification with only a single K\"ahler-modulus, all fields but the universal K\"ahler modulus are stabilized perturbatively. This means that all massless bulk fields have UV-masses $m_{UV}^2/k^2$ that are at most suppressed by the CY-volume which we assume to be only moderately large. Therefore, we may estimate the UV-tails of perturbations by keeping the normalizable mode only for all throat perturbations.\par 
The universal K\"ahler modulus $\rho$ plays a very different role: the volume modulus corresponds to a rescaling of the bulk CY where warping is weak. Regions of strong warping are not affected \cite{Giddings:2005ff}. This means that in the RS1 language the universal K\"ahler modulus is a UV-brane field stabilized with non-perturbative mass scale $m_{\rho}^2\ll k^2$ \cite{Brummer:2005sh}. It couples universally to all UV-perturbations $\delta\phi_{UV}$,
\begin{equation}
\mathcal{L}_{UV}\supset \frac{1}{2}m_{\rho}^2\rho^2+ \Lambda^{3/2}\delta \phi_{UV} \rho+...\, ,
\end{equation}
where the coupling $\Lambda^{3/2}\delta \phi_{UV}\rho$ is perturbative and we have omitted higher order couplings.
Then, a shift $\delta \phi_{UV}$ leads to a shift in $\delta\rho$ of order
\begin{equation}
\delta \rho\sim \frac{\Lambda^{3/2}}{m_{\rho}^2}\delta\phi_{UV}
\end{equation}
Let us work in the regime of moderately large CY-volume, strong warping $e^{\mathcal{A}_0}\ll 1$ and weak non-perturbative stabilization $m_{\rho}^2/k^2\ll 1$. Let us further neglect dependencies on the CY-volume and only retain the dependence on $e^{\mathcal{A}_0}$ and $m_{\rho}$ which we take to be parametrically small. Then,
\begin{equation}
\delta \rho\sim \frac{e^{4\mathcal{A}_0}}{m_{\rho}^2}\delta \phi_{IR}\, .
\end{equation}
For the non-perturbative volume stabilization of KKLT and uplift with an $\overline{D3}$-brane of tension $T_3$ this implies the (parametric) estimate
\begin{equation}
\delta \rho\sim \frac{e^{4\mathcal{A}_0} T_3}{|\langle \lambda \lambda \rangle|^2}\, .
\end{equation}
\subsection{Three-form perturbations}\label{App:sub:RadialProfiles.threeform}
For the three-form field strength $G_3$ the analogous analysis was done in \cite{Baumann:2010sx}. Setting the dilaton constant for simplicity, the linearized three-form equation of motion and Bianchi identity are
\begin{equation}
d(e^{4\mathcal{A}}G^-)=0=dG_3 \, .
\end{equation}
They are solved by a product ansatz
\begin{equation}
G_3=\sum_{i}d\left((a_{i}r^{\delta_{i}-4}+b_{i}r^{-\delta_i})\Omega_2^{(i)}\right)\, ,
\end{equation}
where the $\Omega^{(i)}_2$ are eigen-forms of the Laplace-Beltrami operator $*_5 d$ of $T^{1,1}$,
\begin{equation}
*_5d\Omega_2^{(i)}=i\delta_i \Omega_2^{(i)}\, ,
\end{equation}
with imaginary eigenvalues $i\delta_i$.\par 
In \cite{Baumann:2010sx} it is found that the ISD/IASD pieces are given by
\begin{equation}
G^{+}=G_0^++\frac{i}{2}\sum_{i\neq 0}\left(\frac{a_i}{\delta_i}r^{\delta_i-4}+2b_i r^{-\delta_i}\right)\left(d\Omega_2^{(i)}-\delta_i\frac{dr}{r}\wedge \Omega_2^{(i)}\right)\, ,
\end{equation}
\begin{equation}
G^-=G_0^--i\sum_{i\neq 0}a_i\frac{\delta_i-2}{\delta_i}r^{\delta_i-4}\left(d\Omega_2^{(i)}+\delta_i\frac{dr}{r}\wedge \Omega_2^{(i)}\right)\, ,
\end{equation}
where the zero mode pieces are given by
\begin{equation}
G_0^{\pm}=-2a_0r^{-4}\left(*_5\Omega_2^{(0)}\pm i \frac{dr}{r}\wedge \Omega_2^{(0)}\right)\, .
\end{equation}
Only those eigen two-forms with $\delta_i<2$ give rise to normalizable IASD perturbations. The lowest eigenvalues of the Laplace-Beltrami operator are listed in tables $3$-$5$ of \cite{Baumann:2010sx}. It can be read off that the zero mode falls off least rapidly towards the UV, namely as $r^{-4}$. For ISD perturbations the story is a bit different as also those modes that correspond to the coefficients $b_i$ contribute. Those that feature $\delta_i>2$ are normalizable. The one that falls off least rapidly towards the UV  corresponds to $\delta=3$ and has a radial profile that is proportional to $r^{-3}$ \cite{Baumann:2010sx}. This corresponds to the two-form 
\begin{equation}
\Omega_2^{(\delta=3)}=(g^1+ig^4)\wedge(g^2-ig^3)\, ,
\end{equation}
using the basis of one-forms of $T^{1,1}$ used in \cite{Klebanov:2000hb}. Far from the tip, the KS solution can be viewed as a perturbation of the KT solution and the normalizable $\delta=3$ mode precisely encodes the leading perturbation that is sourced by the deformation at the tip \cite{Loewy:2001pq}. Hence, it is of Hodge type $(2,1)$ and does not enter \eqref{4dpotentialIIB}. As a consequence, the perturbation of $3$-form fluxes on \eqref{4dpotentialIIB} is irrelevant to the order of precision we are interested in, analogous to the scalar case.
\subsection{Metric perturbations}\label{App:sub:RadialProfiles.metric}
In \cite{Gandhi:2011id} the radial scaling of metric modes was derived. After gauge fixing, the only non-vanishing modes are those that correspond to transverse traceless two-tensors of $T^{1,1}$,
\begin{equation}
\delta g_{ij}\sim \sum_{n}\left(a_n r^{\Delta_n(g)-4}+b_n r^{-\Delta_n(g)}\right)Y^{(n)}_{ij}\, ,\quad \Delta_n(g)\equiv 2+\sqrt{\hat{\lambda}_n^2-4}\, ,
\end{equation}
where the $Y^{(n)}_{ij}$ are the transverse traceless eigentensors of the Lichnerowicz operator of $T^{1,1}$ with eigenvalue $\hat{\lambda}^2_n$,
\begin{equation}
\nabla^2 Y^{(i)}_{ij}-2\nabla^k \nabla_{(i}Y^{(i)}_{j)k}=-\hat{\lambda}_n^2Y^{(i)}_{ij}\, .
\end{equation}
In \cite{GMS} it is found that the lowest eigenvalues are $\hat{\lambda}^2=4,5,36-8\sqrt{7},65/4,20,...$ corresponding to
\begin{equation}
\Delta_n(g)=2,3,5.29,11/2,6...\, .
\end{equation}
The dominant normalizable mode thus falls off as $r^{-3}$ and corresponds to the deformation of the conifold. All further perturbations fall off at least as $r^{-5.29}$.
%%%%%%%%%%%%%%%%%%%%%%%%%%%%%%%%%%%%%%%%%%%%%%%%%%%%%%%%%%%%%%%%%%%%%%%%%%

\bigskip
\bibliography{Towards10DdS}
\bibliographystyle{JHEP}
\end{document}